# Dryland evapotranspiration from remote sensing solar-induced chlorophyll fluorescence: constraining an optimal stomatal model within a two-source energy balance model

**Jingyi Bu**[a,b,k], **Guojing Gan**[d], **Jiahao Chen**[a,b], **Yanxin Su**[a,b], **Mengjia Yuan**[a,b], **Yanchun Gao**[a*], **Francisco Domingo**[e], **Mirco Migliavacca**[f,g], **Tarek S. El-Madany**[f], **Pierre Gentine**[h,i,j], **Monica Garcia**[c,k]

[a] Key Laboratory of Water Cycle and Related Land Surface Processes, Institute of Geographical Sciences and Natural Resources Research, Chinese Academy of Sciences, Beijing 100101, China
[b] University of the Chinese Academy of Sciences, Beijing 100049, China
[c] Research Center for the Management of Environmental and Agricultural Risks (CEIGRAM), E.T.S. de Ingeniería Agronómica, Alimentaria y de Biosistemas, Universidad Politécnica de Madrid, P.º de la Senda del Rey, 13, 28040 Madrid, Spain.
[d] Key Laboratory of Watershed Geographic Sciences, Nanjing Institute of Geography and Limnology, Chinese Academy of Sciences, Nanjing 210008, China
[e] Estación Experimental de Zonas Áridas, Consejo Superior de Investigaciones Científicas (CSIC), Ctra. de Sacramento s/n La Cañada de San Urbano, 04120 Almería, Spain
[f] Max Planck Institute for Biogeochemistry, Department Biogeochemical Integration, Jena, Germany
[g] European Commission, Joint Research Centre, Via Fermi 2749, Ispra (VA)
[h] Department of Earth and Environmental Engineering, Columbia University, New York, NY, USA
[i] Department of Earth and Environmental Sciences, Columbia University, New York, NY, USA
[j] Earth Institute, Columbia University, New York, NY, USA
[k] Department of Environmental Engineering, Technical University of Denmark, Lyngby, 2800, Denmark

*Correspondence to **Yanchun Gao**

Contact information:

**Jingyi Bu: bujy.16b@igsnrr.ac.cn**

**Yanchun Gao: gaoyanc@igsnrr.ac.cn**

# Abstract

Evapotranspiration (ET) represents the largest water loss flux in drylands, but ET and its partition into plant transpiration (T) and soil evaporation (E) are poorly quantified, especially at fine temporal scales. Physically-based remote sensing models relying on sensible heat flux estimates, like the two-source energy balance model, could benefit from considering more explicitly the key effect of stomatal regulation on dryland ET.

The objective of this study is to assess the value of solar-induced chlorophyll fluorescence (SIF), a proxy for photosynthesis, to constrain the canopy conductance ($G_c$) of an optimal stomatal model within a two-source energy balance model in drylands. We assessed our ET estimation using in situ eddy covariance GPP as a benchmark, and compared with results from using the Contiguous solar-induced chlorophyll fluorescence (CSIF) remote sensing product instead of GPP, with and without the effect of root-zone soil moisture on the $G_c$.

The estimated ET was robust across four steppes and two tree-grass dryland ecosystem. Comparison of ET simulated against in situ GPP yielded an average $R^2$ of 0.73 (0.86) and RMSE of 0.031 (0.36) mm at half-hourly (daily) timescale. Including explicitly the soil moisture effect on $G_c$, increased the $R^2$ to 0.76 (0.89). For the CSIF model, the average $R^2$ for ET estimates also improved when including the effect of soil moisture: from 0.65 (0.79) to 0.71 (0.84), with RMSE ranging between 0.023 (0.22) and 0.043 (0.54) mm depending on the site.

Our results demonstrate the capacity of SIF to estimate subdaily and daily ET fluxes under very low ET conditions. SIF can provide effective vegetation signals to constrain stomatal conductance and partition ET into T and E in drylands. This approach could be extended for regional estimates using remote sensing SIF estimates such as CSIF, TROPOMI-SIF, or the upcoming FLEX mission, among others.

# 1. Introduction

Drylands cover about 42% of the Earth's surface and feed more than 38% of population around the world (D'Odorico et al., 2019; Reynolds et al., 2007). Due to intensive human activities and climate change, the rate and range of desertification has been reported to increase in some arid/semi-arid areas over the past several decades (Berdugo et al., 2021; Mirzabaev et al., 2019).

Water scarcity, highly variable precipitation and droughts characterize drylands, typically with a short and variable rainy season (Swift, 1993). In those regions, around 90% of annual precipitation returns to the atmosphere as evapotranspiration (ET), but the ET flux is poorly quantified especially at fine temporal scales (Sun et al., 2019). Even more challenging is to quantify the ET partition into plant transpiration (T) and soil evaporation (E) (Cavanaugh et al., 2011; Wang et al., 2012). The T/ET ratio tends to be relatively low (around 0.5) in drylands over the long term compared to wetter climates where can be up to around 0.7 (Schlesinger and Jasechko, 2014; Sun et al., 2019). Stomatal control is a key control in the T/ET ratio that shows large and rapid variations during wetting and drying cycles (Sun et al., 2019). Quantifying dryland ET and its partition accurately at high spatial and temporal resolutions is critical to monitor ecosystem function and develop effective restoration strategies in these ecologically fragile areas (Garcia et al., 2008; Li et al., 2019b).

Remote sensing (RS) techniques can provide critical information correlated with vegetation structure or land surface temperature (LST), indirectly related to water stress or ET (Burchard-Levine et al., 2021a; Garcia et al., 2013; Wang et al., 2012). However, remote sensing based models of surface conductance capable to simulate the wide range of plant physiological responses under water stress are still needed (Berg and McColl, 2021; Li et al., 2015; Scott et al., 2014). Various canopy conductance ($G_c$) models have been implemented within remote sensing-based ET models, based on the upscaling of stomatal models such as the Jarvis model (Jarvis, 1976; Mu et al., 2011), the Ball-Berry model (Ball et al., 1987; Hu et al., 2017), or optimal stomatal models (OSMs) (De Kauwe et al., 2015; Medlyn et al., 2011).

Considering the tight coupling of the water and carbon cycles, where $G_c$ is one of the main joint regulating factors (Gentine et al., 2019), the rate of photosynthesis can be used to constraint $G_c$ (Bonan et al., 2014). Various GPP models have been introduced to constrain $G_c$ and ET (Hu et al., 2013; Ryu et al., 2011; Zhang et al., 2019a). However, their accuracy depends on the validity of model assumptions and the quality of input datasets. In drylands, estimates of vegetation density and the fraction of absorbed photosynthetically active radiation (APAR) tend to be underestimated due to low vegetation signal-to-noise ratios or the high soil background reflectance of sparse vegetation (Fensholt et al., 2004; Garcia and Ustin, 2001; Smith et al., 2019; Yan et al., 2019). This impacts the accuracy of ET, resulting from a lower transpiration area and lower GPP in the case of LUE-based models (Garcia et al., 2013).

Another obstacle to monitoring ET as well as GPP from remote sensing is the decoupling between vegetation functioning and optical reflectance, especially in the case of dryland ecosystems (Smith et al., 2019; Wang et al., 2012). Global retrievals of solar-induced chlorophyll fluorescence (SIF) from GOSAT (Frankenberg et al., 2011), GOME-2 (Joiner et al., 2013), OCO-2 (Frankenberg et al., 2014), TanSat (Du et al., 2018), and TROPOMI (Koehler et al., 2018) provide novel opportunities to monitor photosynthesis at global scales (Mohammed et al., 2019). Emitted from pigment-protein complexes in leaf chlorophyll, SIF is a byproduct of photosynthesis strongly correlated with GPP across biomes at various time scales (Guanter et al., 2014; Li et al., 2018; Liu et al., 2017; Zhang et al., 2018b; Zhang et al., 2019b). Smith et al. (2018) showed that SIF captured GPP dynamics better than vegetation indices or the photochemical reflectivity index (PRI), especially in drylands of Southwestern North America.

Considering the tight linkage between T and photosynthesis (Lu et al., 2018; Maes et al., 2020), the potential of using SIF as a proxy for T has been explored in recent years. Lu et al. (2018) found that the band combinations of canopy SIF emission in near-infrared spectra (720 nm, 740 nm, and 760 nm) were more sensitive to T than single-band SIF in temperate forests, and the relationship between SIF and T did not change diurnally. Pagán et al. (2019) showed that the SIF/PAR ratio was able to capture the effects of environmental stress on T. Maes et al. (2020) confirmed the relatively high correlation (r=0.76) between T and GOME-2/OCO-2 SIF across global FLUXNET sites. Shan et al. (2019) found that $G_c$ and SIF presented similar patterns diurnally and seasonally and developed a linear relationship between SIF and $G_c$ to compute T. Further, Shan et al. (2021) derived a SIF-based $G_c$ model using an optimal stomatal model and a strong relationship between SIF as $G_c \cdot VPD^{0.5}$. These latter results demonstrated that the SIF-based semi-mechanical model offered more predictive power in T estimation than a simple linear model over forests, croplands, and grasslands. Feng et al. (2021) put forward two mechanistic methods to predict T using canopy SIF from the scanning spectrometer based on the WUE and $G_c$ respectively, at both half-hourly and daily scales. Zhou et al. (2022) further combined SIF and meteorological data to estimate ET based on Fick's law and an optimal stomatal behavior.

Damm et al. (2021) suggested that SIF should be incorporated in models combined with various Earth observation data to enhance ET estimation. Though the ET or T estimated from SIF and $G_c$ has made progress, most of those efforts have concentrated on the big-leaf Penman-Monteith (PM) ET model (Damm et al., 2021; Feng et al., 2021; Lu et al., 2018; Shan et al., 2019; Shan et al., 2021). Two source energy balance (TSEB) models (Kustas and Norman, 1999a; Norman et al., 1995), separating energy and land surface temperature (LST) into the canopy and soil layers, can better capture the dynamics of heat fluxes well under non-homogeneous surface conditions than one source energy balance model (Anderson et al., 2007; Burchard-Levine et al., 2020). However, TSEB models are sensitive to errors in LST and its partitioning between the soil and canopy. To address this and improve the biophysical regulation of the canopy module, Gan and Gao (2015) introduced a $G_c$ model into a TSEB model (Leuning et al., 2008) named $G_c$-TSEB. Bu et al. (2021) demonstrated that the $G_c$-TSEB model coupled with an upscaled optimal stomatal conductance model (OSM) (Medlyn et al., 2011) instead of Leuning's original $G_c$ model can improve ET estimation and partitioning, particularly in drylands and croplands. However, the shortcoming for global and regional applications of the $G_c$-TSEB approach is the requirement of GPP as input. Since SIF is regarded as a proxy of GPP at ecosystem scale, it would be possible to constrain the canopy conductance within the $G_c$-TSEB model to improve drylands ET estimates at regional scales.

Coarse spatial resolution and sparse temporal sampling have been a barrier to using RS SIF application (Mohammed et al., 2019). Produced using a machine learning algorithm, a contiguous solar-induced chlorophyll fluorescence (CSIF) (Zhang et al., 2018a) using OCO-2 SIF and MODIS data, with high spatiotemporal resolution (4-day and 0.05°) and long time span (2001-2020), has been applied successfully in global and regional monitoring of ecosystem production (Liu et al., 2020), carbon uptake (Zhang et al., 2020) and plant phenology (Meng et al., 2021).

The main objective of this study is to assess the capacity of Earth Observations of SIF to estimate ET and its partitioning between T and E by constraining the conductance of the $G_c$-TSEB model over six drylands flux sites representative of natural grasslands and savanna biomes. To achieve this goal, we (1) simulated $G_c$ using in-situ EC GPP as a benchmark; (2) simulated $G_c$ with an empirical relationship of SIF with GPP (3) explored the role of root-zone soil water content and its regulation on surface conductance (canopy and soil) and ET modeling.

## 2. Materials and Methods

Some parts of this section are removed from the journal submission version while it goes through the peer review.

2.1 Study areas

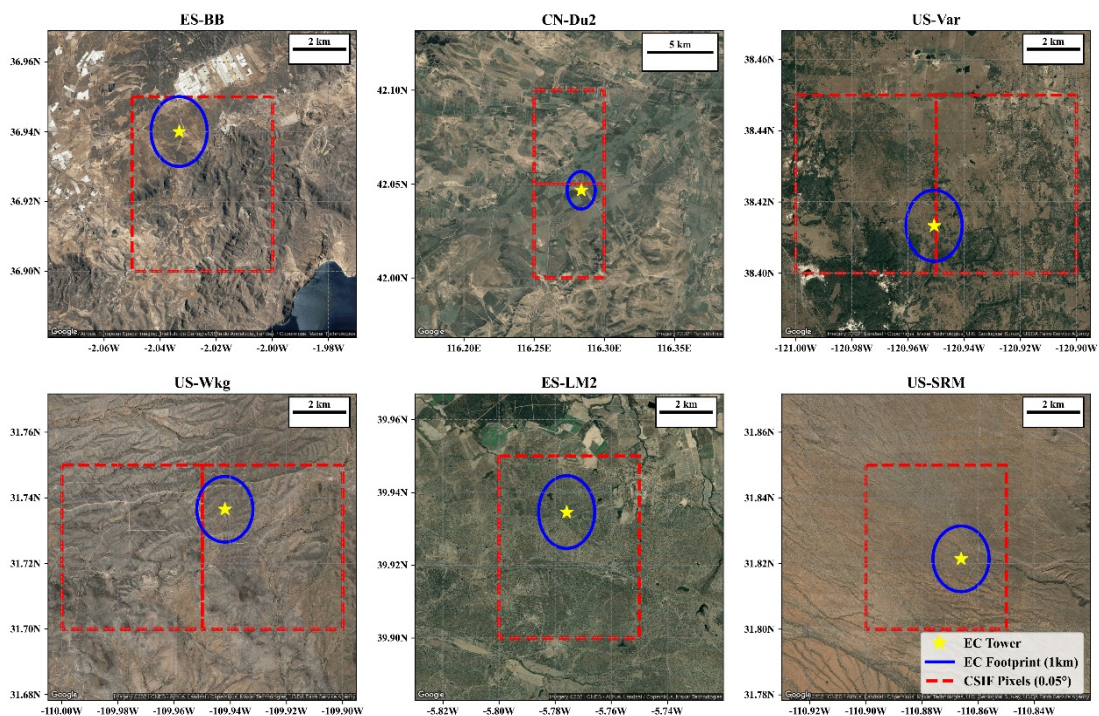

Fig.1 Views of land cover conditions within the footprint of flux towers (usually less than 1 km) and the pixel size of CSIF (0.05°) for six sites

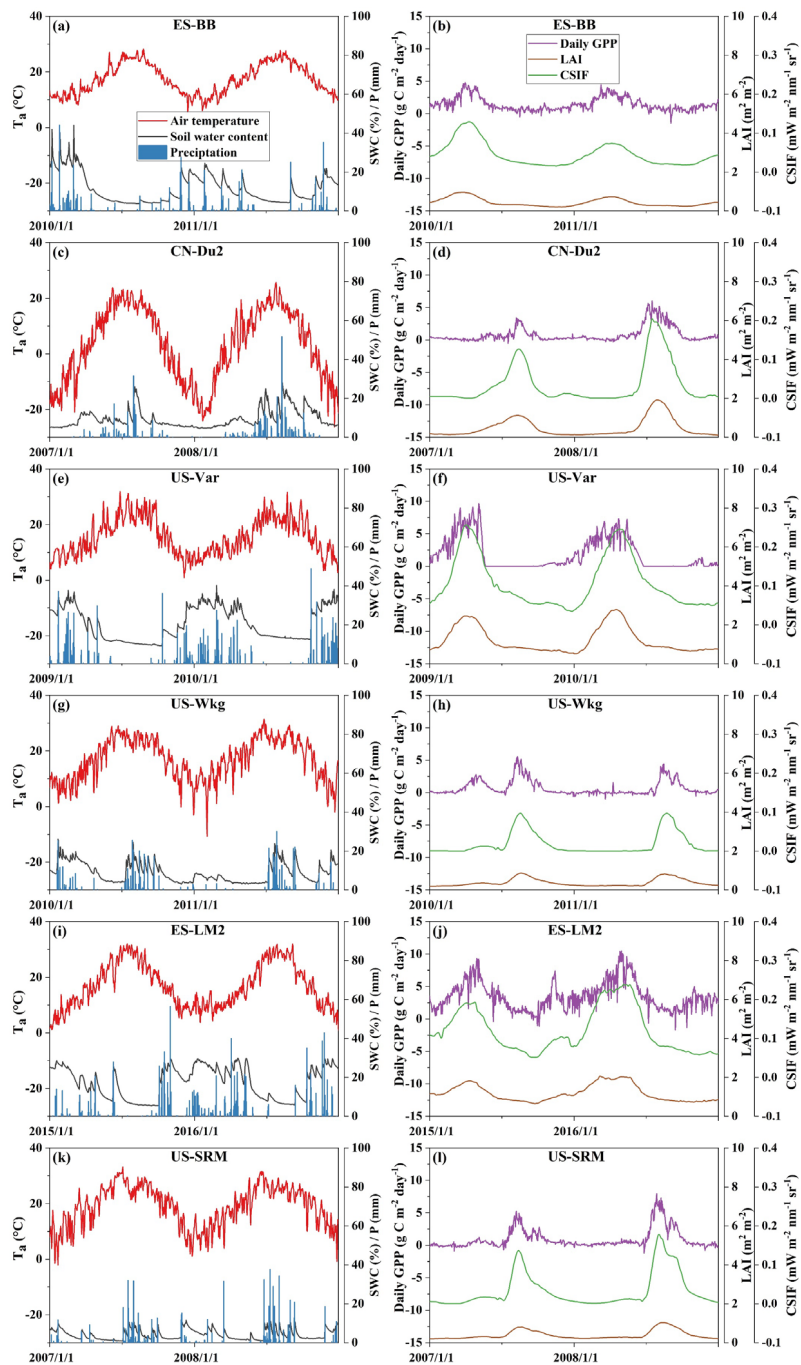

Fig.2 Variation of air temperature ($T_a$), precipitation, root-zone SWC, daily GPP, CSIF, and MODIS LAI of six sites during the study period. Values for ES-BB are shown in (a, b), CN-Du2 in (c, d), US-Var in (e, f), US-Wkg in (g, h), ES-LM2 in (i, j), and US-SRM in (k, l)

Six sites (four grassland sites and two savanna sites) located in the Mediterranean or other semi-arid regions were used to evaluate the $G_c$-TSEB. Detailed site descriptions are shown in

Table 1. The land cover and meteorological conditions at each site see Fig.1 and Fig.2.

**Grassland sites**

The ES-BB site (Balsa Blanca, 36.94°N, 2.03°W) (Garcia et al., 2013; López-Ballesteros et al., 2017) is a permanent tussock grassland in Cabo de Gata Natural Park, Spain, with semi-arid Mediterranean climate. Most of the precipitation falls in autumn and winter with a mean annual precipitation of 375 mm and mean annual temperature of 18.1°C (Morillas et al., 2013a). The dominant vegetation is *Stipa tenacissima L*, with a canopy height of 0.7 m. The soil texture class is sandy loam.

The CN-Du2 site (Duolun Restoration Ecology Experimentation and Demonstration Station, 42.05°N, 116.28°E) (Chen et al., 2009; Miao et al., 2009) is in the southwest of Duolun County, Inner Mongolia of China. The site is categorized as a semi-arid continental temperate climate, with mean annual temperature of 3.3 °C and mean annual precipitation of 399 mm. 88% of rainfall events occur in the growing season from May to September. The steppe around the EC tower is dominated by *S. krylovii*, *A. frigida*, *Agropyron cristatum*, *Cleistogenes squarrosa*, and *Leymus chinensis*. The canopy height of vegetation is between 30 and 40 cm. The Luvic Kastanozems, with 76.8% sand, 16.7% silt, and 6.5% clay, is the main soil coverage of the site.

The US-Var site (Vaira Ranch, 38.41° N; 120.95° W) (Baldocchi et al., 2004; Ma et al., 2007) is located at the footsteps of Sierra Nevada Mountains, California. The site is a semi-arid Mediterranean climate with mild/warm winter while hot/dry summer. The mean annual temperature is 16.6°C and the mean annual precipitation is 546 mm. Most precipitation concentrates from October to next May. The *Brachypodium distachyon*, *Hypochaeris glabra*,

*Bromus madritensis*, and *Cynosurus echinatus* cover most of the steppe. The dominant soil type is Chromic Luvisols, containing high silt (57%), medium sand (30%), and low clay (13%).

The US-Wkg site (Walnut Gulch Kendall Grasslands, 31.74°N, 109.94°W) (Cavanaugh et al., 2011; Scott et al., 2010) is located at USDA Walnut Gulch Experimental Watershed of southeastern Arizona. This site presents a semiarid climate within the North American monsoonal regime, with mean annual precipitation of 407 mm and mean annual temperature of 15.64°C. It is characterized by two growing periods, from June to September in summer and in the Spring from February to April. Summertime, when about 60% of rain in a year occurs, is the dominant productive season. The native vegetation used to be dominated by C4 *bunchgrasses* (*Bouteloua eriopoda*) but has been replaced by Lehmann lovegrass (*Eragrostis lehmanniana*) since 2006. The soils are characterized as coarse loamy.

**Savanna sites**

The ES-LM2 site (Majadas de Tietar South, 39.93°N, 5.78°W) (El-Madany et al., 2018; El-Madany et al., 2021) is a typical tree-grass ecosystem (TGE) situated in the Mediterranean climate zones of central Spain. The mean annual precipitation is 636 mm, with 85% of this coming in the growing season, from October to April. The mean temperature is 16.7°C. The site is randomly spread by evergreen broadleaf Holm Oak (*Quercus ilex L.*), taking up around 20% of the total land area, and the herbaceous layer is dominated by grasses, forbs, and legumes. The canopy heights of oak and herbaceous are 8 and 0.1-0.3m respectively, with a weighted average height of 2 m (Burchard-Levine et al., 2020). Soils around ES-LM2 are Abruptic Luvisol mixed with sand.

The US-SRM site (Santa Rita mesquite savanna site, 31.82°N, 110.87°W) (Cavanaugh et al., 2011; Scott, 2010) lies in the Santa Rita Experimental Range, Arizona in the USA. The mean annual temperature is 17.92°C with a maximum temperature exceeding 35°C in June. The annual mean precipitation is 377 mm. Similar to US-Wkg, about 50% of precipitation is brought by the summer monsoon, which falls between July and September. The native vegetation consisted of annual herbs (*Digitaria californica* Benth. and *Muhlenbergia porteri*). But with the invasion and spreading of alien mesquite (*Prosopis velutina* Woot.), the semi-arid steppe has evolved into a woody savanna over the past hundred years. The mesquite, taking up about 35% of the site, has a canopy height ranging between 0.25 and 6 m, with a mean value of 2.5 m. The soils are characterized as loamy sand.

Table 1 A description of selected sites

| Station ID | Country | Lat. | Lon. | Ele (m) | $T_a$ (°C) | P (mm) | $H_c^a$ (m) | $ET^b$ (mm) | $\beta^c$ | Years | IGBP | Soil texture | Depth of root-zone SWC (m) | $SWC_f$ (%) | $SWC_w$ (%) | References |
|---|---|---|---|---|---|---|---|---|---|---|---|---|---|---|---|---|
| ES-BB | Spain | 36.94 | -2.03 | 196 | 18.1 | 375 | 0.7 | 141 | 2.55 | 2010-2011 | GRA | Sandy loam | 0.4 | 14 | 4 | (Garcia et al., 2013; López-Ballesteros et al., 2017) |
| CN-Du2 | China | 42.05 | 116.28 | 1324 | 3.3 | 399 | 0.4 | 219 | 1.64 | 2007-2008 | GRA | Sandy loam | 0.1 | 26 | 4 | (Chen et al., 2009; Miao et al., 2009) |
| US-Var | America | 38.41 | -120.95 | 129 | 15.8 | 559 | 1 | 340 | 0.97 | 2009-2010 | GRA | Silt | 0.1 | 31 | 9 | (Baldocchi et al., 2004; Ma et al., 2007) |
| US-Wkg | America | 31.74 | -109.94 | 1531 | 15.6 | 407 | 0.5 | 255 | 2.79 | 2010-2011 | GRA | Loam | 0.05 | 18 | 2 | (Cavanaugh et al., 2011; Scott et al., 2010) |
| ES-LM2 | Spain | 39.93 | -5.78 | 264 | 16.7 | 636 | 2 | 513 | 1.47 | 2015-2016 | SAV | Sandy loam | 0.3 | 20 | 5 | (El-Madany et al., 2018; El-Madany et al., 2021) |
| US-SRM | America | 31.82 | -110.87 | 1120 | 17.9 | 380 | 2.5 | 302 | 2.90 | 2007-2008 | SAV | Loamy sand | 0.1 | 11 | 0.5 | (Cavanaugh et al., 2011; Scott, 2010) |

a Canopy height
b Annual ET
c Bowen ratio

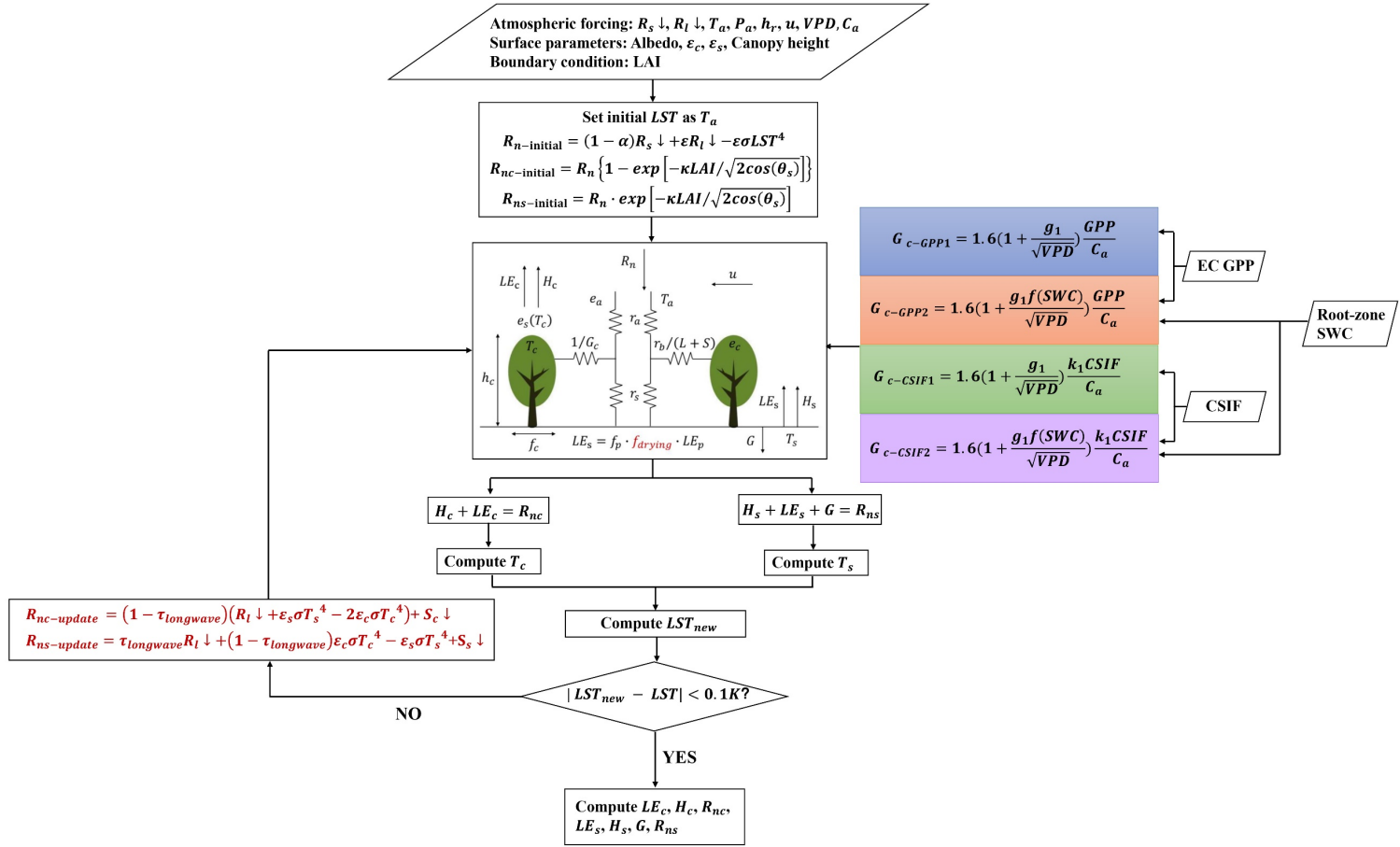

Fig.3 The flowchart of G$_c$-TSEB, revised from Gan and Gao (2015) The red texts refer to the improvements that are different from the original G$_c$-TSEB.

## 2.2 Model overview

**Net radiation**

In the G$_c$-TSEB model, the net radiation ($R_n$, W m$^{-2}$) is the sum of incoming and outgoing shortwave and longwave radiation (Gan and Gao, 2015):

$$R_n = (1 - \alpha)R_s \downarrow + \varepsilon_s R_l \downarrow - \varepsilon_s \sigma LST^4 \tag{1}$$

where $R_s \downarrow$ and $R_l \downarrow$ are downward shortwave and longwave radiation (W m$^{-2}$) respectively. $\sigma$ is the Stefan-Boltzmann constant (5.67 × 10$^{-8}$ W m$^{-2}$ K$^{-4}$). $LST$ is the land surface temperature (K). $\alpha$ and $\varepsilon_s$ are the surface albedo and emissivity, respectively. Following previous studies of the G$_c$-TSEB model, the $R_n$ is partitioned into canopy ($R_{nc}$, W m$^{-2}$) and soil fractions ($R_{ns}$, W m$^{-2}$) following a layered approach according to Beer's Law (Kustas et al., 1998):

$$R_{nc} = R_n \left\{ 1 - exp\left[ -\kappa LAI / \sqrt{2\cos(\theta_s)} \right] \right\} \tag{2}$$

$$R_{ns} = R_n \cdot exp\left[ -\kappa LAI / \sqrt{2\cos(\theta_s)} \right] \tag{3}$$

where $\kappa$ is the radiative transfer extinction coefficient, and $\theta_s$ is the solar zenith angle (rad). As high soil temperature may influence the contribution of soil thermal radiation to net radiation, this method produces significant systematic errors when applied in sparse vegetation regions (Kustas and Norman, 1999b). Considering the different radiative transfer response for canopy and soil in response to visible and near-infrared (NIR) wavelengths of spectrum, Kustas and Norman (1999a) updated the $R_n$ estimation algorithm as follows and that was implemented in our study:

$$R_{nc} = (1 - \tau_{longwave})(R_l \downarrow + \varepsilon_s \sigma T_s^4 - \varepsilon_c \sigma T_c^4) + S_c \tag{4}$$

$$R_{ns} = \tau_{longwave} R_l \downarrow + (1 - \tau_{longwave})\varepsilon_c \sigma T_c^4 - \varepsilon_s \sigma T_s^4 + S_s \tag{5}$$

$$\tau_{longwave} = exp(-k_L LAI) \tag{6}$$

where $\tau_{longwave}$ is the longwave transmittance. $\varepsilon_c$ and $\varepsilon_s$ are canopy and soil emissivity, set as 0.98 and 0.95 respectively in this study. $T_c$ and $T_s$ are canopy and soil temperature (K) respectively. $k_L$ is the longwave radiation extinction coefficient, set as 0.95. $S_c$ and $S_s$ are the net shortwave radiation (W m$^{-2}$) reaching canopy and soil that can be derived taking into account direct and diffuse light fractions by:

$$S_c = (1-\tau_{bV})(1-\alpha_{cbV})S_b F_V + (1-\tau_{bN})(1-\alpha_{cbN})S_b F_N + \\ (1-\tau_{dV})(1-\alpha_{cdV})S_d F_V + (1-\tau_{dN})(1-\alpha_{cdN})S_d F_N \tag{7}$$

$$S_s = \tau_{bV}(1-\alpha_{sV})S_b F_V + \tau_{bN}(1-\alpha_{sN})S_b F_N + \tau_{dV}(1-\alpha_{sV})S_d F_V + \\ \tau_{dN}(1-\alpha_{sN})S_d F_N \tag{8}$$

where $\tau_{b*}$ and $\tau_{d*}$ are direct beam (subscript b) and diffuse (subscript d) canopy transmittance respectively. Subscript hereafter (indicated by a "*") refers to visible (subscript V) and NIR (subscript N) regions of the radiation. $\alpha_{cb*}$ and $\alpha_{cd*}$ are direct beam and diffuse canopy albedo. $\alpha_{sV}$ ($\alpha_{sN}$) is the soil albedo in visible (NIR) wavelength, set as 0.15 (0.25) in grassland and 0.07 (0.19) in savanna (He et al., 2019; Kool et al., 2021). $S_b$ and $S_d$ are the direct beam and diffuse shortwave radiation. $F_V$ and $F_N$ are the fraction of total visible and NIR radiation.

This method coupled with the TSEB model (for more details about the model see Kustas and Norman (1999a)) was successfully applied in cropland (Song et al., 2016b) and dryland (Morillas et al., 2013a). It is worth noting that, $T_c$ and $T_s$ are key state variables to estimate net longwave radiation. If $T_c$ or $T_s$ are lacking, the $R_{nc}$ and $R_{ns}$ cannot be determined, which is the first step of the model iteration. Unlike in the original TSEB version (Kustas and Norman, 1999a), $LST$ in G$_c$-TSEB is a model output rather than an input (Fig. 3). To obtain initial $T_c$ and $T_s$, we still apply Eq. (1)-(3) to retrieve $R_{nc}$ and $R_{ns}$ as initial conditions for model calculation. After the first iteration, $T_c$ and $T_s$ are obtained from Eqs. (4)-(8), instead of Eqs. (1)-(3), in the following iterations.

**Energy balance for the canopy layer**

The energy absorbed by the canopy is dissipated as canopy sensible heat flux ($H_c$, W m$^{-2}$)

and latent heat flux ($LE_c$, W m$^{-2}$):

$$H_c + LE_c = R_{nc} \quad (9)$$

where $H_c$ is directly proportional to the difference between air temperature ($T_a$, K) and $T_c$ (Gan et al., 2019):

$$H_c = \rho c_p \frac{T_c - T_a}{r_a + r_b} \quad (10)$$

where $\rho$ and $c_p$ represent air density (1.25 kg m$^{-3}$) and the specific heat of air at the constant pressure (1005 J kg$^{-1}$ K$^{-1}$), respectively. The aerodynamic resistance ($r_a$, s m$^{-1}$) and leaf boundary layer resistance ($r_b$, s m$^{-1}$) were computed according to Gan and Gao (2015).

Different from the original TSEB model which used a Priestley-Taylor (PT) equation (Kustas and Norman, 1999a), $LE_c$ in G$_c$-TSEB is derived according to the mass transfer equation as the difference of the atmosphere vapor pressure ($e_a$, kPa) and canopy vapor pressure ($e_s(T_c)$, kPa):

$$LE_c = \frac{\rho c_p}{\gamma} \frac{e_s(T_c) - e_a}{r_a + r_c} \quad (11)$$

where $\gamma$ is the psychrometric constant (0.066 kPa K$^{-1}$). $r_c$ represents the canopy resistance (s m$^{-1}$), and is the reciprocal of canopy conductance ($1/G_c$), which will be further discussed in Section 2.3.

**Energy balance for the soil layer**

In the soil, in addition to sensible heat flux ($H_s$, W m$^{-2}$) and latent heat flux ($LE_s$, W m$^{-2}$), $R_{ns}$ can be dissipated as soil heat flux ($G$, W m$^{-2}$).

$$H_s + LE_s + G = R_{ns} \quad (12)$$

$H_s$ and $G$ can be expressed as follows:

$$H_s = \rho c_p \frac{T_s - T_a}{r_a + r_s} \quad (13)$$

$$G = \alpha_G R_{ns} \tag{14}$$

where $\alpha_G$ represents the proportion of $R_{ns}$ dissipated as G and needs to be calibrated. $r_s$ is the canopy resistance (s m$^{-1}$) between the soil and the canopy displacement height and is calculated according to Zeng et al. (2005):

$$r_s = 1/(c_t u_*) \tag{15}$$

where $c_t$ is the turbulent transfer coefficient and $u_*$ is the friction velocity reaching the canopy (m s$^{-1}$). The value of $c_t$ depends on the interpolation between a bare soil turbulent transfer coefficient ($c_{t,bare}$) and full canopy turbulent transfer coefficient ($c_{t,dense}$, set as 0.04).

$$c_t = c_{t,bare} w_t + c_{t,dense}(1 - w_t) \tag{16}$$

$$c_{t,bare} = \frac{k}{m}\left(\frac{z_{0m,g} u_*}{v}\right)^{-0.45} \tag{17}$$

where $k$ is the von Karman constant (0.4). Constant $m$ is set as 0.13. $z_{0m,g}$ is the momentum roughness length for bare soil, set as 0.01m. $v$ is the kinematic viscosity of air (1.5×10$^{-5}$ m$^2$ s$^{-1}$). The empirical interpolation weight ($w_t$) was originally formulated as a function of LAI and stem area index (S) as $e^{-(LAI+S)}$ (Zeng et al., 2005). Gan and Gao (2015) assumed that the canopy height ($h_c$) also has a certain effect on the interpolation and revised the formulation as $e^{-(L+\Omega h_c)}$ which we adopted in our study $\Omega$ is a combination coefficient of canopy height and stem area index and needs to be calibrated.

In the G$_c$-TSEB, the calculation of $LE_s$ relies on the complementary concept between actual and potential evaporation (Gan et al., 2019; Gao et al., 2016):

$$LE_s = LE_{s\_PM} \cdot f_p \tag{18}$$

where $LE_{s\_PM}$ is the potential evaporation derived from the Penman-Monteith equation:

$$LE_{s\_PM} = \frac{\Delta(R_{ns} - G) + \rho c_p (e_s(T_a) - e_a)/(r_a + r_s)}{\Delta + \gamma} \tag{19}$$

where $e_s(T_a)$ is the saturated vapor pressure of the air at given $T_a$ (kPa) and $\Delta$ is the slope of saturation-to-vapor pressure curve (kPa K$^{-1}$).

$f_p$ represents the relative soil evaporative fraction and is expressed as an exponential function of a surface dryness index $\frac{LE_{p\_mt}}{LE_{p\_mt}+R_{ns}-G}$ (Gao et al., 2016; Granger and Gray, 1989):

$$f_p = \frac{1}{1 + a_{PM} \cdot exp\left(b_{PM} \cdot \frac{LE_{p\_mt}}{LE_{p\_mt} + R_{ns} - G}\right)} \tag{20}$$

$$LE_{p\_mt} = \frac{\rho c_p}{\gamma} \frac{e_s(T_S) - e_a}{r_a + r_s} \tag{21}$$

where $e_s(T_S)$ is the vapor pressure near the soil surface (kPa). $a_{PM}$ and $b_{PM}$ are empirical parameters that need to be calibrated.

Considering that soil evaporation is highly sensitive to the soil moisture and precipitation pulses over drylands (Cavanaugh et al., 2011), we introduced in this study a smoothing factor, $f_{drying}$ (Zhang et al., 2010), to better account for soil evaporation following rain pulses:

$$LE_s = LE_{s\_PM} \cdot f_p \cdot f_{drying} \tag{22}$$

$$f_{drying} = min\left(0.01, \frac{\sum_{n=1}^{N} P_n}{\sum_{n=1}^{N} E_{s,n}}, 1\right) \tag{23}$$

where $P_n$ and $E_{s,n}$ are the total precipitation (mm) and available energy for soil evaporation (mm) on a specific day (n). $f_{drying}$ is the ratio of accumulated precipitation and soil available energy over N days. N was taken as 32 following Zhang et al. (2019a). To avoid degenerating $f_{drying}$ to 0 during the long drying season, we set minimum value of $f_{drying}$ as 0.01 in this study.

**Calculation procedures and generation of LST**

The workflow of the G$_c$-TSEB model used is illustrated in Fig. 1. As the surface energy balance is dependent on the LST, the purpose of model iterations is to solve for the aggregated

LST and its components, $T_c$ and $T_s$. Based on the $T_c$ and $T_s$, the surface energy fluxes, LE and H, can be further obtained following the explicit energy balance equations.

a) Step 1: the initial LST in Eq. (1) is set as the air temperature, which implies the evapotranspiration will be maximum and H will be null. Then the initial $R_{nc}$ and $R_{ns}$ can be obtained from Eq. (2) and (3). Using the Newton-Raphson iteration approach, the initial $T_c$ and $T_s$ can be derived from Eq. (9)-(11) and (12)-(23), respectively.

b) Step 2: After the initial $T_c$ and $T_s$ are determined, the $R_n$ calculation algorithm is updated according as Eq. (4)-(8). The iteration is repeated to calculate $T_c$ and $T_s$ from which LST is computed as follows:

$$LST_{new} = [f_c T_c + (1-f_c)T_s^4]^{\frac{1}{4}} \qquad (24)$$

$$f_c = 1 - \exp(-0.5 LAI) \qquad (25)$$

where $f_c$ represents the vegetation fraction.

c) Step 3: Finally, the loop terminates when the condition ($|LST_{new} - LST_{old}| < 0.1K$) is satisfied. The final canopy and soil energy fluxes are then retrieved based on estimated $T_c$ and $T_s$ from Eq (9)-(23).

2.3 Settings for different $G_c$ scenarios

**GPP-based $G_c$ model (benchmark)**

In this study, an optimal leaf stomatal conductance ($g_s$, mol m$^{-2}$ s$^{-1}$) model (OSM) based on an optimal stomatal conductance behavior (Medlyn et al., 2011) was applied in the canopy module of the G$_c$-TSEB. The OSM assumes that (1) the stomatal opening is controlled by ribulose1,5-bisphosphate regeneration and (2) the atmospheric $CO_2$ concentration is much larger than the $CO_2$ compensation point. The quadratic equation is solved for g$_s$ as a function of intercellular CO2 concentration, net assimilation rate ($A_n$, μmol m$^{-2}$ s$^{-1}$), CO2 concentration at leaf surface ($C_s$, μmol mol$^{-1}$), and vapor pressure deficit (VPD) at leaf surface ($D_s$, kPa) (Medlyn et al., 2011):

$$g_s = 1.6\left(1 + \frac{g_1}{\sqrt{D_s}}\right)\frac{A_n}{C_s} \tag{26}$$

where $g_1$(kPa$^{0.5}$) is the stomatal slope parameter that is proportional to the marginal water cost of assimilation and the CO2 compensation point. At the ecosystem level, $G_c$ can be derived from Eq. (26), taking GPP, atmospheric VPD ($D_a$), and CO2 concentration ($C_a$) for $A_n$, $D_s$, and $C_s$ (Medlyn et al., 2017):

$$G_c = 1.6\left(1 + \frac{g_1}{\sqrt{D_a}}\right)\frac{GPP}{C_a} \tag{27}$$

To account for the effect of soil water stress on $G_c$, $f(SWC)$, a function of soil moisture variation was introduced to OSM.

$$G_c = 1.6\left(1 + \frac{g_1 f(SWC)}{\sqrt{D_a}}\right)\frac{GPP}{C_a} \tag{28}$$

**SIF-based $G_c$ model**

To test the potential of SIF (W m$^{-2}$ μm$^{-1}$ sr$^{-1}$) to constrain $G_c$ in G$_c$-TSEB, GPP was estimated as a function of SIF. Both GPP and SIF can be linked to LUE-based models (Guanter et al., 2014):

$$GPP = APAR \times \text{LUE}_p \tag{29}$$

$$SIF = APAR \times \text{LUE}_F \times \text{f}_{esc} \tag{30}$$

where $APAR$ (W·m$^{-2}$) is the absorbed photosynthetically active radiation (PAR) by leaf photosynthetic pigments, $\text{LUE}_p$ and $\text{LUE}_F$ are the light use efficiency of APAR utilized in photosynthesis and SIF yield respectively, and $\text{f}_{esc}$ is the fraction of emitted fluorescence that escapes from the plant canopy (sr$^{-1}$). Combining Eq. (29) and (30), GPP can be derived from SIF as follows:

$$GPP = SIF \times \frac{LUE_p}{LUE_F \times f_{esc}} \quad (31)$$

Because the ratio $\frac{LUE_p}{LUE_F \times f_{esc}}$ can be regarded as a relatively constant at the ecosystem scale and over weekly time scales (Guanter et al., 2014; Li et al., 2018; Liu et al., 2017; Zhang et al., 2019b), Eq. (29) can be written as a proportional relationship between SIF and GPP:

$$GPP = SIF \times k_1 \quad (32)$$

where $k_1$ is a parameter that needs to be calibrated per ecosystem. Substituting Eq. (31) into Eq. (27) and Eq. (28), the $G_c$ model based on SIF without and with SWC can be obtained as follows:

$$G_c = 1.6 \left(1 + \frac{g_1}{\sqrt{D_a}}\right) \frac{k_1 SIF}{C_a} \quad (33)$$

$$G_c = 1.6 \left(1 + \frac{g_1 f(SWC)}{\sqrt{D_a}}\right) \frac{k_1 SIF}{C_a} \quad (34)$$

2.4 Model calibration and validation

There are six model parameters that need to behave been calibrated, in our case we did it against EC observations, but they could be also calibrated by satellite LST as in Bu et al. (2021). The a priori ranges of the parameters were taken as follows: $g_1$: 0-14 kPa$^{0.5}$, $k_1$: 5-100, $a_{PM}$: 0.01-0.1, $b_{PM}$: 6.5-15, $\Omega$: -0.5-0.5, $\alpha_G$: 0.03-0.35, derived from the literatures (Feng et al., 2021; Gan and Gao, 2015; Gan et al., 2019; Medlyn et al., 2017). Because the $g_1$ and $k_1$ are constant for a specific ecosystem type, these two parameters were calibrated in the scenarios without soil moisture content. The parameters were estimated using a Monte Carlo Markov Chain method (5000 replicates in each case). The objective function was the sum of the Root-Mean-Squared-Error (RMSE) for the estimated LE and H. 70% of days in the study period were selected randomly for parameters calibration, while the rest is for validation. Half-hourly meteorological and flux data satisfying PAR>0, $R_n$>0, $T_a$>0, LE>0, H>0, and GPP>0 were used to calibrate and validate the model. Days without data around noon and with rainfall were removed.

2.5 LST validation

To evaluate the modeled LST, an estimated LST was obtained based on the longwave outgoing radiation sensor, available at US-Var, US-Wkg, ES-LM2, and US-SRM, following Burchard-Levine et al. (2020):

$$LST = \frac{R_l \uparrow - (1-\varepsilon_s)R_l \downarrow}{\sigma \varepsilon_s} \qquad (36)$$

where $R_l \uparrow$ represent outgoing longwave radiation.

2.6 Sensitivity Analysis

To quantify the model sensitivity due to the uncertainties in the parameters and their range, a variance-based global Sensitivity Analysis (SA) following Sobol′ (2001) was applied. Sobol' SA can calculate the main effect of one input variable on output variance individually (First-order indices, $S_i$) and also determine the contribution when considering the interactions with the other model parameters (Total-effect index, $S_{Ti}$). For details on the calculation processes we refer the reader to (Burchard-Levine et al., 2020; Nossent et al., 2011; Sobol′, 2001). The objective function was the sum of RMSE between the estimated LE and H from the model and EC observations.

The six model parameters: $g_1$, $k_1$, $a_{PM}, b_{PM}, \Omega, \alpha_G$, which need to be calibrated in $G_c$-TSEB were assessed using Sobol' SA for the different scenarios, and the ranges of parameters (Section 2.3). Perturbations to monthly time series for the selected model input variables: the CSIF (or the EC GPP in GPP scenarios), LAI, and SWC, were set to ± 10% around the monthly average value based on the study by Garcia et al. (2013).

2.7 Comparing of ET partitioning

To validate the reliability of the estimated T and E at the six dryland sites, we partitioned the observed ET during the growing season (GPP > max(GPP)*0.1) into T and E based on the ratio of apparent underlying WUE ($uWUE_a$) and potential uWUE ($uWUE_p$) (Zhou et al., 2016), and compared it against the simulations using either CSIF or EC GPP.  The comparison between the ET partitioning results of Gc-TSEB and uWUE method, and precipitation (P) was mainly based on correlation analysis.

$$uWUE = \frac{GPP \cdot \sqrt{D_a}}{ET} \tag{37}$$

$$T = ET \cdot \frac{uWUE_a}{uWUE_p} \tag{38}$$

2.8 Meteorological and flux tower data

The half-hourly meteorological data, air temperature, VPD, incoming shortwave and longwave radiation, PAR, wind speed, air pressure, atmospheric $CO_2$ concentration, and EC GPP, were used to drive the $G_c$-TSEB model. To correct the EC energy imbalances, the observed LE and H were post-processed using the 'Bowen ratio forced closure' method (Twine et al., 2000). Because a four-component radiometer was not installed at the ES-BB site during the study period, we applied observed incoming PAR, air temperature, air pressure, and effective atmospheric emissivity to calculate $R_s \downarrow$ and $R_l \downarrow$ at this site, according to Sugita and Brutsaert (1993). Root-zone SWC was measured at depths varying between 5 and 40 cm, depending on the site (Table 1).

2.9 Remote sensing forcing data

The LAI applied in this study was obtained from the MODIS product MCD15A3H.V006 with a spatial resolution of 500 m and temporal resolution of 4-day (Myneni et al., 2015). The albedo, from the MODIS product MCD43A3.V006, has a 500 m pixel size and daily interval (Schaaf and Wang, 2015). The MODIS LAI and albedo were resampled into 1 km grid resolution, by nearest-neighbor interpolation, and 1-day, by the Savitzky-Golay filter (Savitzky and Golay, 1964) and linear interpolation.

The SIF data came from the MODIS-based contiguous solar-induced fluorescence (CSIF) datasets at 0.05° spatial resolution and 4-day temporal resolution (Zhang et al., 2018a). Using CSIF is constructed by filling spatial and temporal gaps of the Orbiting Carbon Observatory-2 (OCO-2) SIF data at 757 nm using neural networks machine learning algorithm and MODIS surface reflectance (MCD43C4.V006) to reproduce OCO2-SIF, with high quality. The time series at each site were smoothed by SG filter and resampled to daily time scale by linear interpolation.

## 3. Results

3.1 Performance of $G_c$-TSEB with GPP from eddy covariance

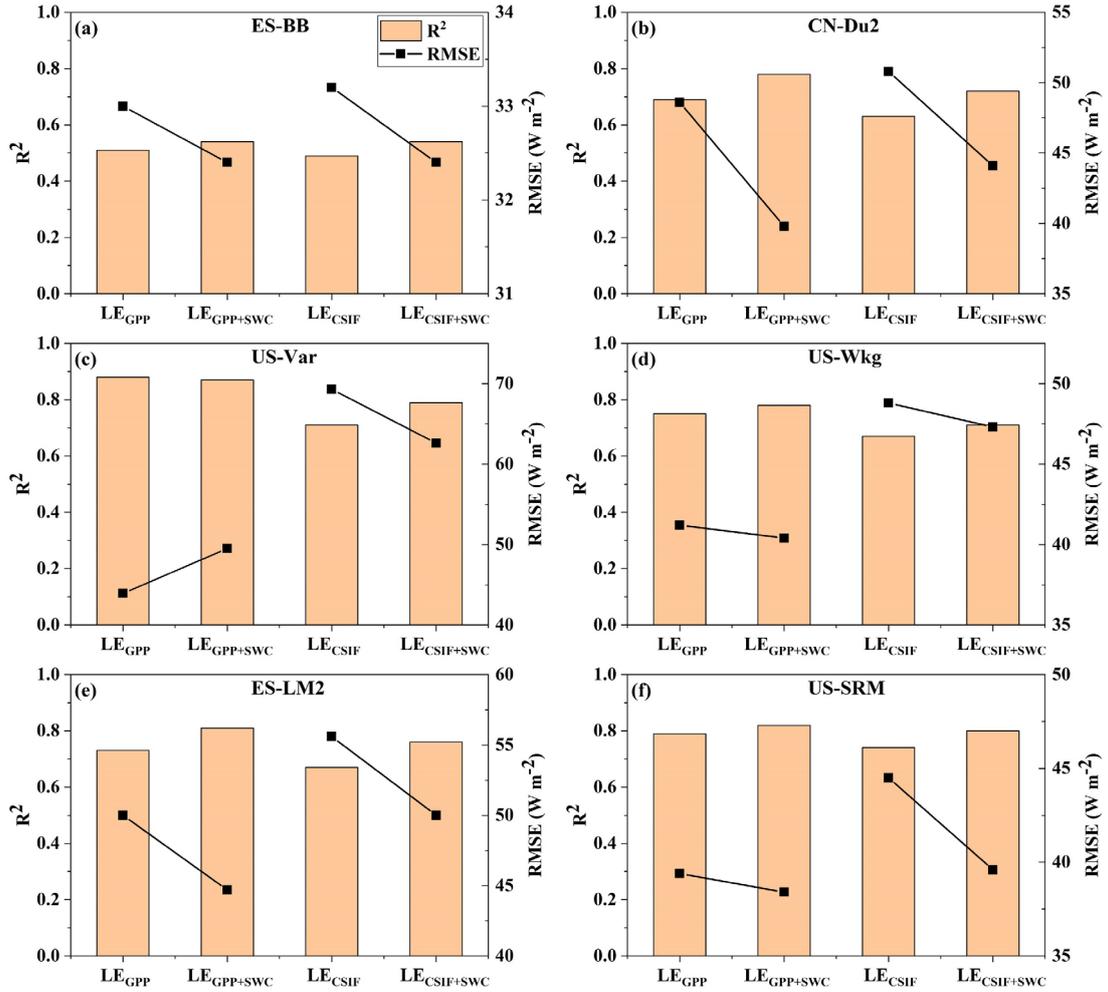

Fig.4 Accuracy metrics of half-hourly LE for EC GPP and SIF scenarios (without and with root-zone SWC) in validation groups at six sites.

The scatter plots and statistical metrics of half-hourly LE simulation by $G_c$-TSEB in conjunction with OSM are shown in Fig. 4 and 5. Across most dryland sites, except ES-BB, the model performed well on half-hourly LE, with RMSE between 37.5 (39.4) and 51.7 (50.0) W m$^{-2}$, and $R^2$ between 0.72 (0.69) and 0.88 (0.88) in calibration (validation) sets. The addition of root-zone soil moisture control on $G_c$ improved the half-hourly LE simulation for most sites but to different degrees (Fig. 4). The most obvious improvement in the model performance including SWC occurred at CN-Du2 (GRA) and ES-LM2 (SAV), with respective $R^2$ increase by 13% and 11%, and RMSE decreased by 8.8 and 5.3 W m$^{-2}$ in the validation sets. Though root-zone soil moisture did not improve the LE modeling in the site of US-Var (GRA), with lowest Bowen ratio, the satisfactory results (mean $R^2$ > 0.87 and mean relative RMSE < 33%) were obtained in two scenarios, both without and with SWC, suggesting reduced control of root zone soil moisture on surface conductance at US-Var than at the other sites.

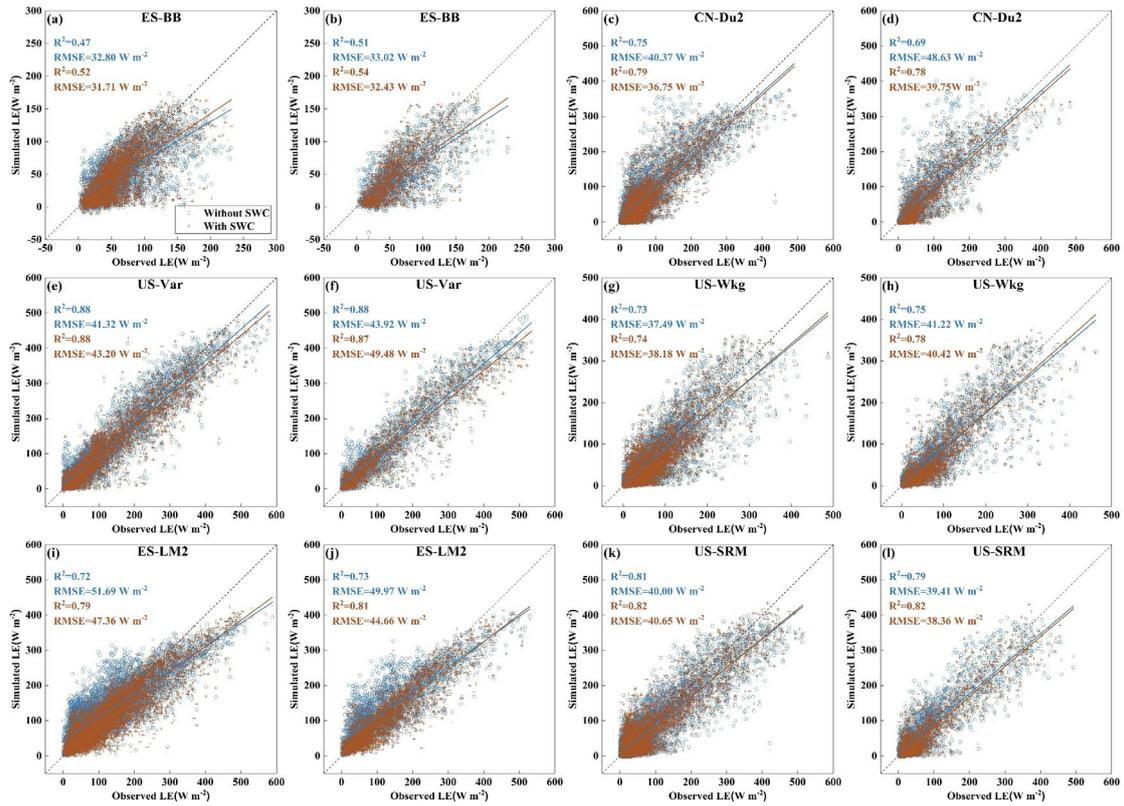

Fig.5 Scatter plots of estimated half-hourly LE against observed LE for GPP scenarios (GPP from Eddy Covariance) at six sites (Blue points: the simulations without SWC, Brown points: the simulations with SWC). (a, c, e, g, i, k) refer to calibration groups, and (b, d, f, h, j, l) refer to validation groups.

The daily statistics and time series in Fig. 6 show that the $G_c$-TSEB in EC GPP scenarios without and with SWC control was able to capture variations of daily ET at most dryland sites, with respective $R^2$ of 0.86 and 0.89, and RMSE of 0.36 and 0.35 mm day$^{-1}$. The model performances at the daily scale were superior to those at the half-hourly scale.

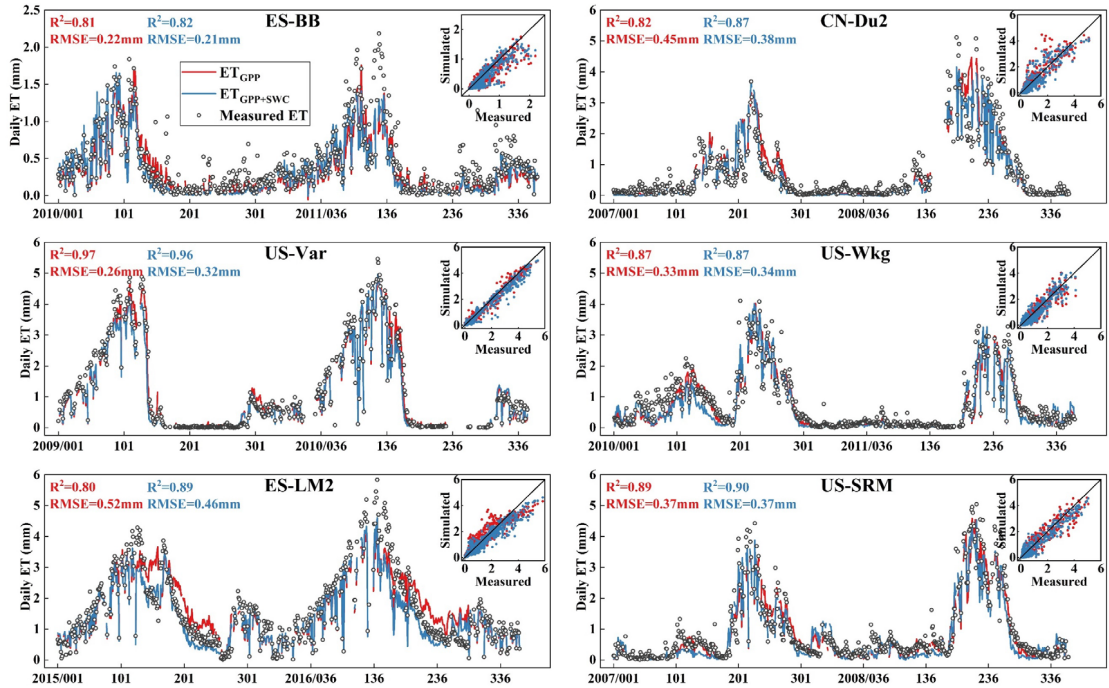

Fig.6 Time series and scatter plots of estimated daily ET against observed ET for GPP scenarios (GPP from Eddy Covariance) at six sites

3.2 Performance of $G_c$-TSEB with Sun Induced Fluorescence (CSIF) datasets

Using a CSIF-based $G_c$ model without SWC control (Fig. 7), $R^2$ values for half-hourly LE ranged from 0.45 (0.49) at ES-BB to 0.75 (0.74) at US-SRM and RMSE values ranged from 32.45 (33.21) W m$^{-2}$ at ES-BB to 63.53 (69.28) W m$^{-2}$ at US-Var for the calibration (validation) datasets, slightly less accurate in general than for the GPP scenarios. When introducing root-zone SWC into $G_c$, the simulations improved for all six sites, with mean $R^2$ increasing from 0.65 to 0.71 and mean RMSE decreasing from 49.40 W m$^{-2}$ to 45.56 W m$^{-2}$. Simulations using CSIF, both without and with additional SWC information, are slightly less accurate in general than for the direct use of EC GPP in the model (Fig. 4), except, surprisingly, for the drier sites with very low ET values: ES-BB and US-SRM. At the sites where H tends to be larger than LE, the model also performed pretty well for the dominant flux (Table S2). Worth noting, when using CSIF instead of EC GPP, the highest loss of accuracy for simulation happened at US-Var.

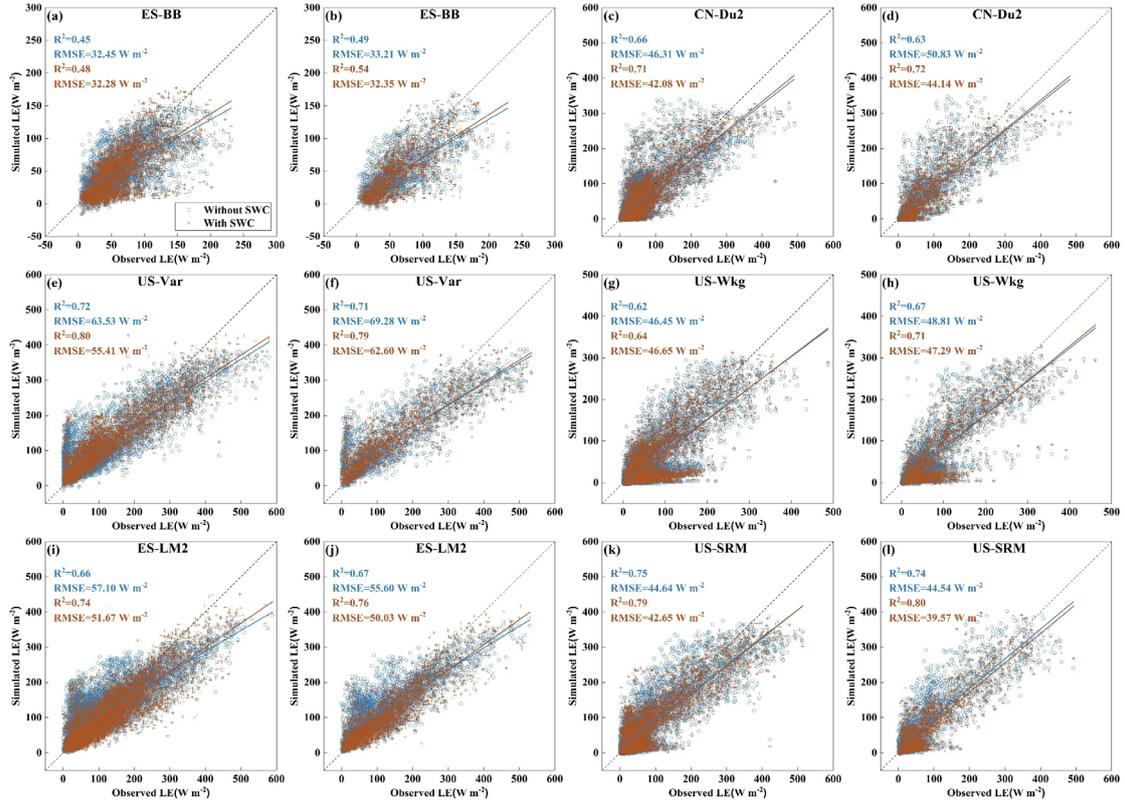

Fig.7 Scatter plots of estimated half-hourly LE against observed LE for CSIF scenarios at six sites (Blue point: the simulations without SWC, Brown point: the simulations with SWC). (a, c, e, g, i, k) refer to calibration groups, and (b, d, f, h, j, l) refer to validation groups.

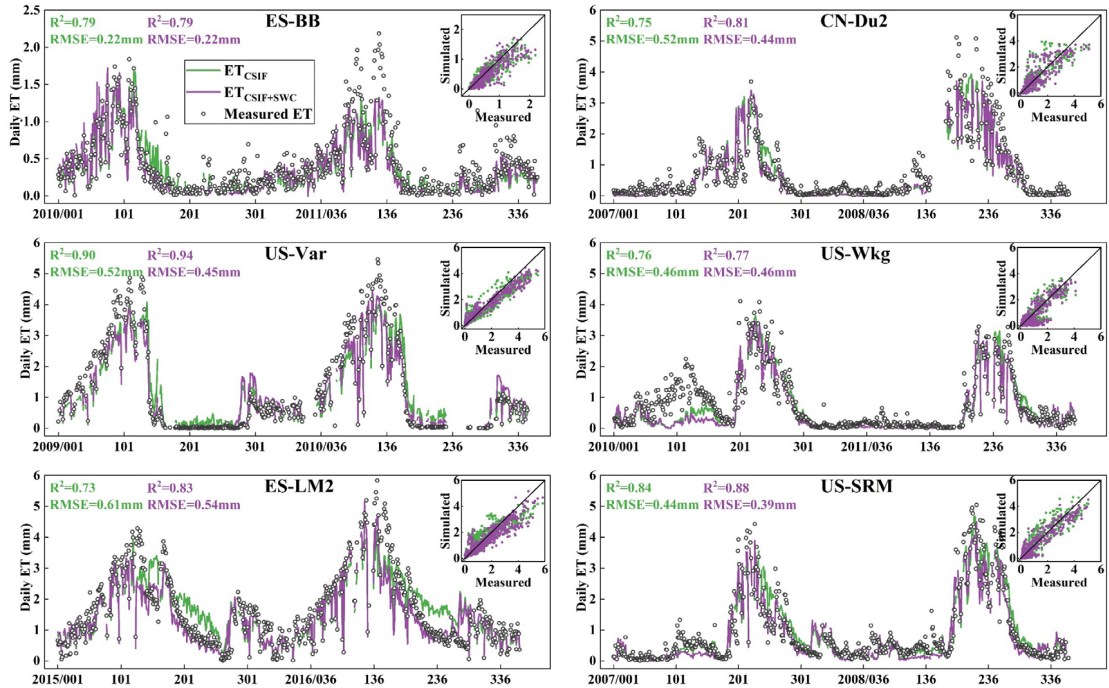

Fig.8 Time series and scatter plots of estimated daily ET against observed ET for CSIF scenarios at six sites

For the daily scale, Fig. 8, $G_c$-TSEB combined with CSIF and root-zone SWC had better performance and accuracy on ET simulation (mean $R^2$=0.84, mean RMSE=0.42 mm day$^{-1}$) than that without SWC (mean $R^2$=0.79, mean RMSE=0.46 mm day$^{-1}$). Moreover, with the additional SWC control on $G_c$, the results using CSIF in CN-Du2 ($R^2$=0.81, RMSE=0.44 mm day$^{-1}$), ES-LM2 ($R^2$=0.83, RMSE=0.54 mm day$^{-1}$), and US-SRM ($R^2$=0.88, RMSE=0.39 mm day$^{-1}$), were in the range to those using EC GPP without SWC (Fig. 6). Sometimes $R^2$ for ET is very high due to the high correlation between $R_n$ and ET, with a marked seasonal cycle. To avoid such spurious correlations, we calculated the evaporation fraction (EF) that normalizes the seasonal changes in $R_n$. The results (Fig. 9) show that the estimated EF was consistent with the observations. SWC improved the EF estimation at most sites, particularly in combination with CSIF, with average $R^2$ from 0.55 to 0.61.

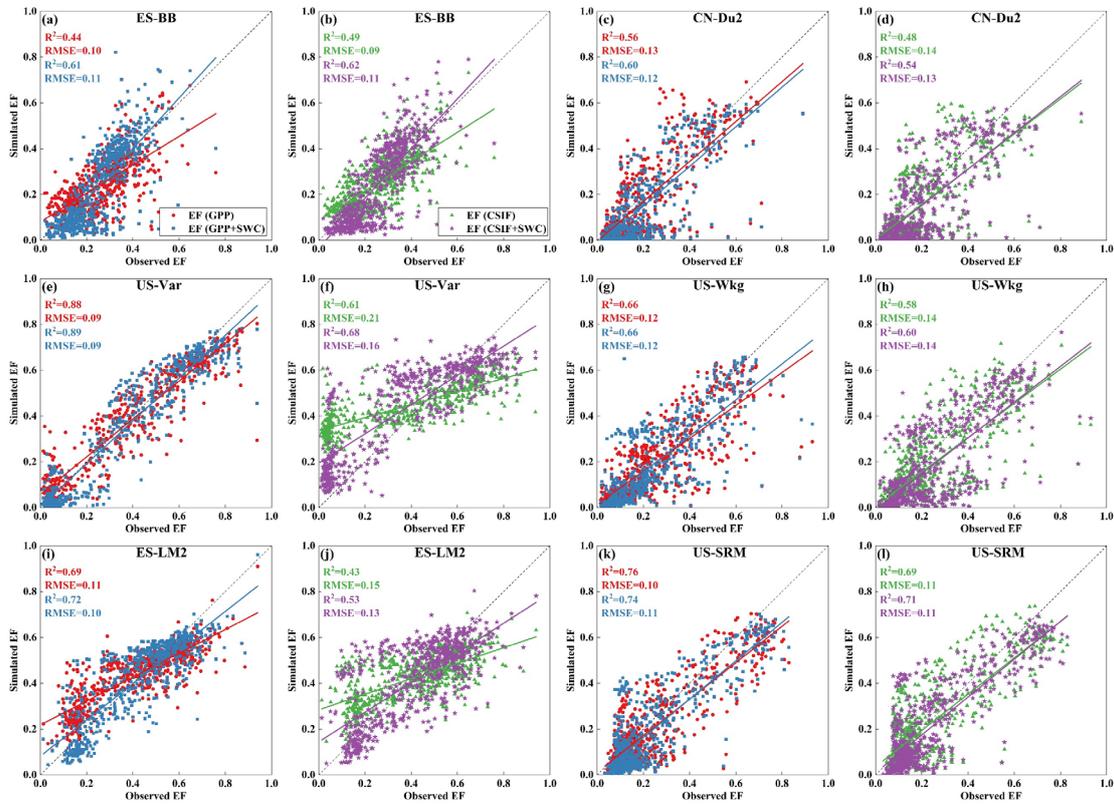

Fig.9 Scatter plots of estimated daily evaporation fraction (EF=LE/($R_n$-G)) against observed EF at six sites (Red point: the simulations using GPP, Blue rectangle: the simulations using

GPP and SWC, Green triangle: the simulations using CSIF, Purple star: the simulations using CSIF and SWC). (a, c, e, g, i, k) refer to the simulations using GPP, and (b, d, f, h, j, l) refer to the simulations using CSIF. CN-Du2 didn't have G observations, so we calculated LE/$R_n$ as EF.

3.3 Evaluation of energy budgets and LST simulations

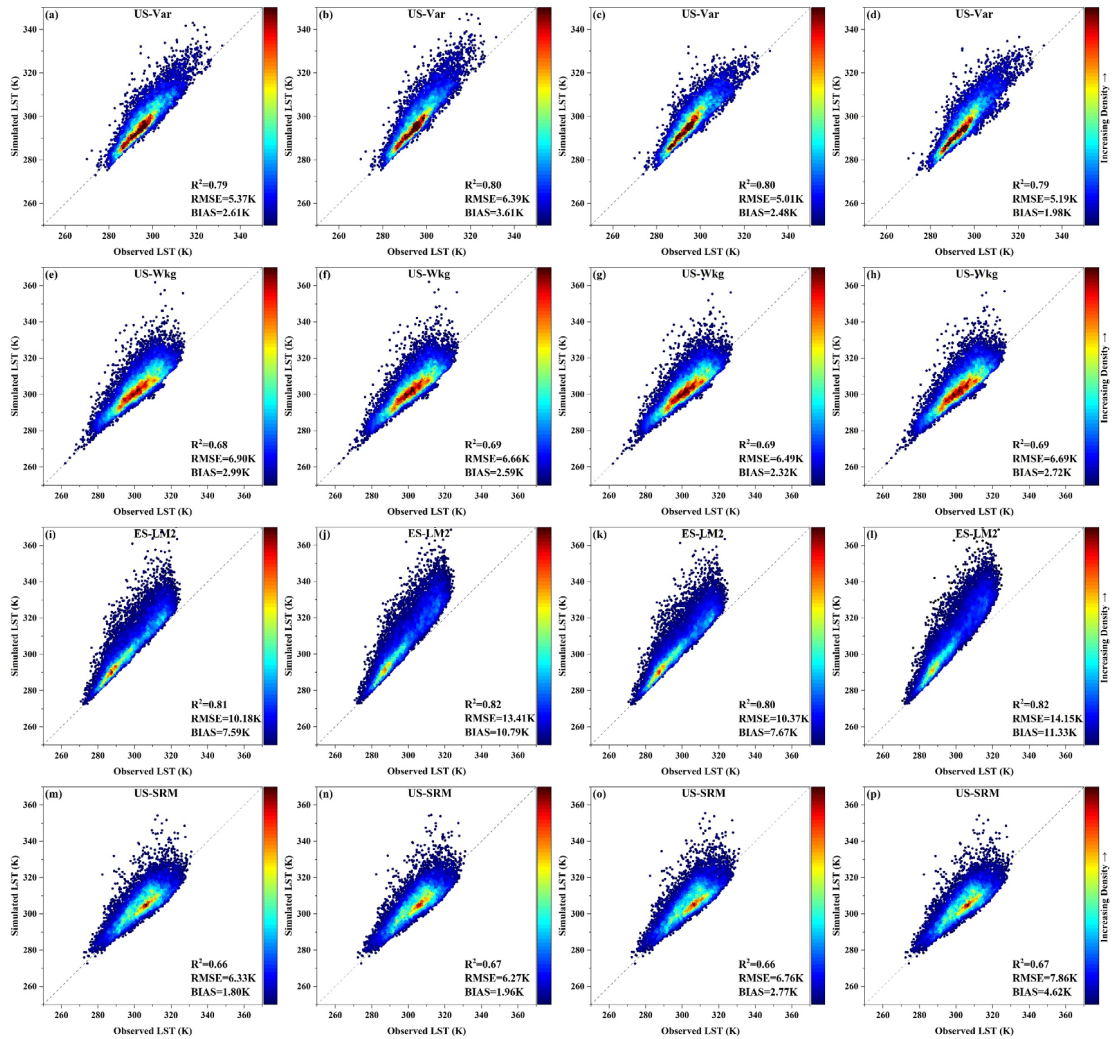

Fig.10 Scatter plots of estimated half-hourly LST against observed LST (From in-situ measurements of Longwave radiation) at six sites. GPP scenarios without SWC are shown in (a, e, i, m), GPP scenarios with SWC in (b, f, j, n), CSIF scenarios without SWC are shown in (c, g, k, o), and CSIF scenarios with SWC are shown in (d, h, l, p).

For the $G_c$-TSEB model, the accuracy of the energy fluxes was determined through the evaluation of the estimated $R_n$ and LST. The model produced satisfactory estimations for $R_n$ (Table S3), except for the ES-BB site. The principal reason is that the separate measured shortwave and longwave radiation fluxes were not available at ES-BB. The large bias (> 30%) of $R_n$ led to noticeable deviation for LE (bias > 20%, Table S1) at this site with relatively low corresponding ET values. As shown in the density scatter plots and statistics results (Fig. 10), the model had good performance on LST estimation at four sites, with both relative RMSE and BIAS taking up less than 25% of the value of ($LST_{max}$-$LST_{min}$). There were no significant differences between simulations using EC GPP- or CSIF-based estimates of $G_c$, but the additional information from root-zone SWC did not improve the LST estimation (beyond the information already present in GPP). Though SWC control was effective in ET estimation of ES-LM2 (SAV), the RMSE and BIAS for LST increased from 10.18 (10.37) and 7.59 (7.67) to 13.41 (14.15) and 10.79 (11.37) K in the EC GPP (CSIF) group when adding SWC constraints.

3.4 Sensitivity analysis for parameters

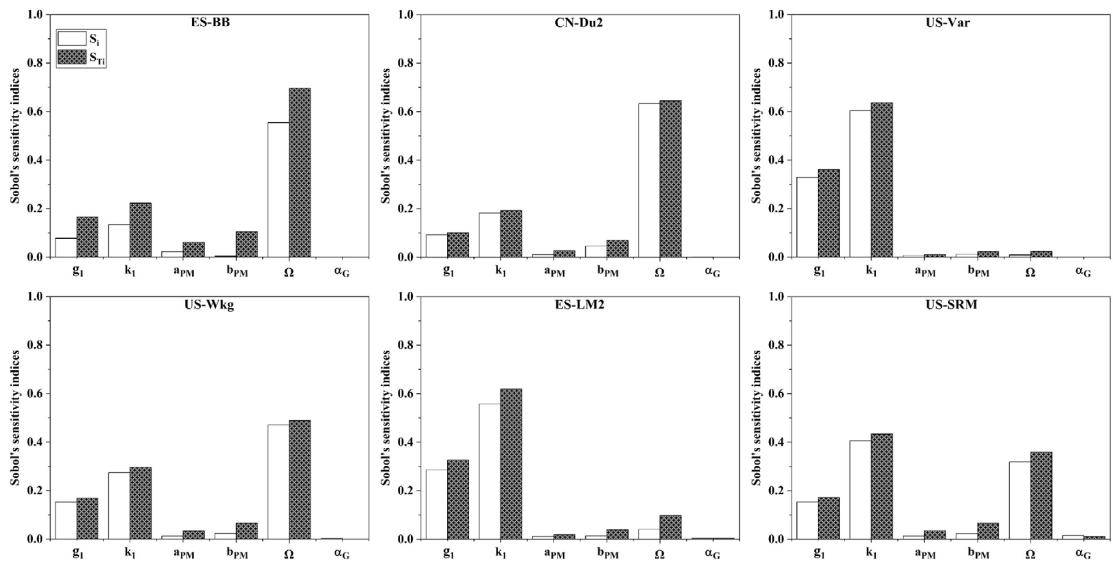

Fig.11 The Sobol' sensitivity indices for heat fluxes (LE and H) from $G_c$-TSEB owing to uncertainty in model parameters. $S_i$ = First-order indices, $S_{Ti}$ = Total-effect indices.

As shown in Fig. 11, parameters related to canopy conductance ($g_1$ and $k_1$) and under-canopy resistance (Ω) had more significant impacts on heat fluxes estimation than the other parameters at most sites. For the relatively more humid sites (US-Var and ES-LM2), $k_1$ significantly affected model simulation with high sensitivities ($S_i > 0.5$ and $S_{Ti} > 0.6$), followed by $g_1$, while the values for other parameters were less than 0.1. For the drier sites (ES-BB, CN-Du2, US-Wkg, and US-SRM), where annual precipitation is less than 500mm, the $G_c$-TSEB model showed high sensitivity to Ω. Parameters related to soil evaporation ($a_{PM}$ and $b_{PM}$) and soil heat fluxes ($\alpha_G$) had a small effect on the model performance. In addition, the value of $S_i$ closely approximated that of $S_{Ti}$ for most parameters, indicating that the interaction between parameters did not affect the simulation of LE and H, and the model was robust.

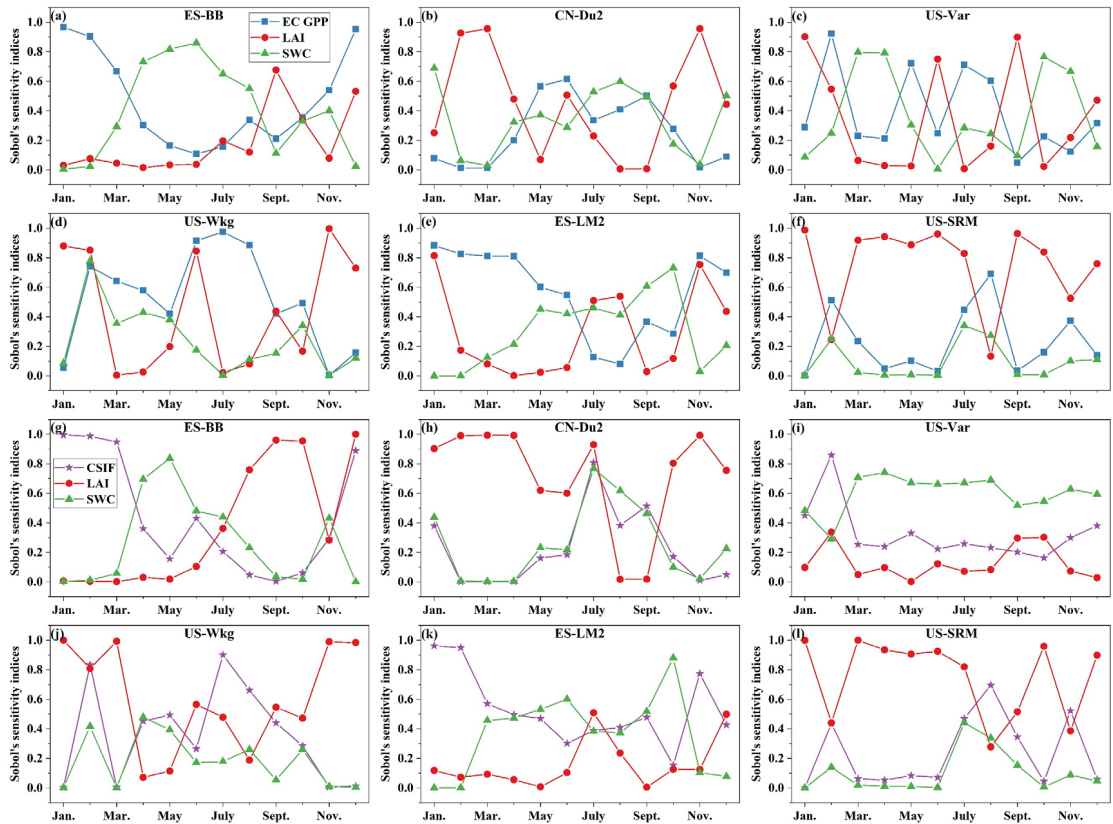

Fig.12 The monthly variations of Sobol' sensitivity indices for heat fluxes (LE and H) from $G_c$-TSEB owing to uncertainty in model key inputs, EC GPP, CSIF, LAI, and SWC.

The monthly Sobol's total sensitivity indices for four model input variables (EC GPP, CSIF,

LAI, and root-zone SWC) are presented in Fig. 12. In general, the SA temporal dynamics of the ET model with CSIF were roughly identical to those with EC GPP. The $S_{Ti}$ of four variables demonstrated a clear seasonal pattern. In the growing period with high ET, the GPP, and CSIF, played a key role in LE and H estimation ($S_{Ti} > 0.5$). But in the non-growing season, the model was more sensitive to LAI, as would be expected due to phenological changes. During the germinating or greening (depending on life form) and senescence of vegetation, the simulation of LE and H was more sensitive to the variation of root-zone SWC.

3.5 ET partitioning by Gc-TSEB with GPP and CSIF datasets

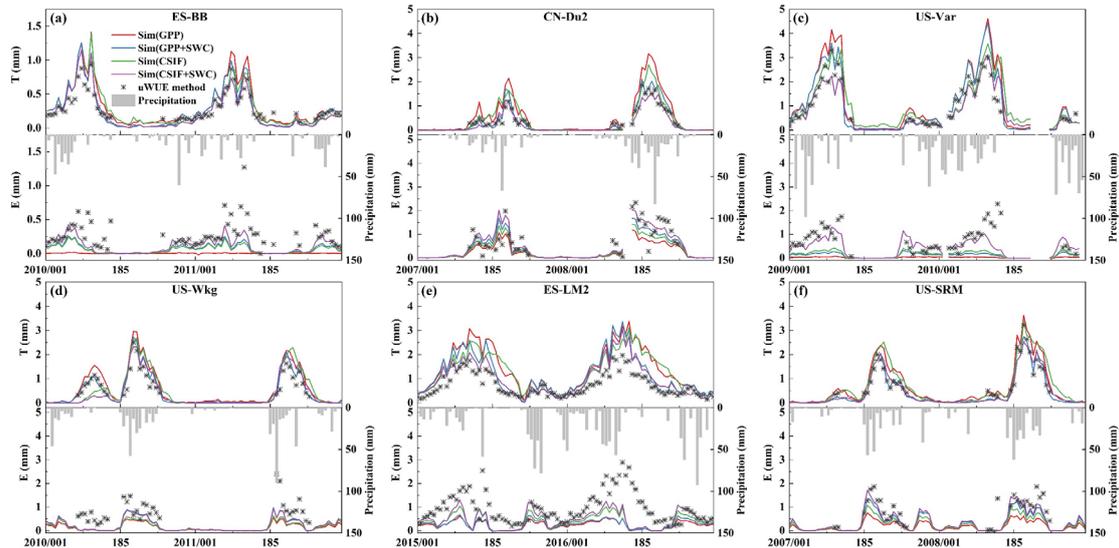

Fig.13 8-day time series of estimated canopy transpiration (T) and soil evaporation (E) against ET partitioning by uWUE method

Overall, the ET partitioning using CSIF was similar to one using GPP based on EC data (Fig. 13 and Table 2). The simulated T and E presented obvious seasonal variations following plant phenology and precipitation events (Fig 13), agreeing with the estimated T:ET partitioning from the uWUE method. As shown in Table 2, the high correlation ($R^2 > 0.5$) indicated a strong relationship between the simulated T by $G_c$-TSEB and $T_{uWUE}$ at most sites.

When adding root-zone SWC information in $G_c$, the estimated T was closer to the observation-based $T_{uWUE}$, particularly in combination with CSIF, with average $R^2$ from 0.74 to 0.81. The additional root-zone SWC control improved the model's capacity of capturing the rapid decline of T during the senescence of vegetation following dry periods (Fig. 13). This improvement was quite clear at ES-BB (Day 133-169 in 2010), CN-Du2 (Day 169-257 in 2007 and Day 185–257 in 2008), ES-LM2 (Day 145-257 in 2015 and Day 177-289 in 2016), and US-SRM (Day 217-289 in 2007 and Day 217–273 in 2008), in which SWC seemed to play a significant role in improving the ET estimation and reducing the T component.

Although high correlations between estimated T and $T_{uWUE}$ were evidenced, the estimated E and $E_{uWUE}$ exhibited a weak relationship (Table 2 and Fig. 13). However, the $R^2$ between E estimated by $G_c$-TSEB and precipitation (mean $R^2$ = 0.24) was higher than that between $E_{uWUE}$ and P (mean $R^2$ = 0.10) at most sites, except US-Wkg and US-Var (Table 2).

Table 2 Determination coefficient ($R^2$) between the ET partitioning results of $G_c$-TSEB (subscript: sim) and uWUE method (subscript: uWUE), and precipitation (P).

| Sites | Simulations | $T_{uWUE}$ & $T_{sim}$ | $E_{uWUE}$ & $E_{sim}$ | P & $E_{sim}$ | P & $E_{uWUE}$ |
|---|---|---|---|---|---|
| ES-BB | Sim(GPP) | 0.96*** | 0.053 | 0.20*** | |
| | Sim(GPP+SWC) | 0.94*** | 0.21*** | 0.12** | 0.0004 |
| | Sim(CSIF) | 0.88*** | 0.18*** | 0.11** | |
| | Sim(CSIF+SWC) | 0.90*** | 0.21*** | 0.12** | |
| CN-Du2 | Sim(GPP) | 0.94*** | 0.50*** | 0.12* | |
| | Sim(GPP+SWC) | 0.90*** | 0.45*** | 0.14* | 0.048 |
| | Sim(CSIF) | 0.79*** | 0.49*** | 0.12* | |
| | Sim(CSIF+SWC) | 0.85*** | 0.46*** | 0.14* | |
| US-Var | Sim(GPP) | 0.88*** | 0.0016 | 0.00090 | |
| | Sim(GPP+SWC) | 0.92*** | 0.036 | 0.032 | 0.026 |
| | Sim(CSIF) | 0.77*** | 0.12* | 0.036 | |
| | Sim(CSIF+SWC) | 0.81*** | 0.16** | 0.04 | |
| US-Wkg | Sim(GPP) | 0.92*** | 0.40*** | 0.22** | |
| | Sim(GPP+SWC) | 0.90*** | 0.41*** | 0.23** | 0.44*** |
| | Sim(CSIF) | 0.52*** | 0.42*** | 0.24** | |

|  | Sim(CSIF+SWC) | 0.56*** | 0.42*** | 0.25** |  |
| --- | --- | --- | --- | --- | --- |
| ES-LM2 | Sim(GPP) | 0.88*** | 0.16*** | 0.16*** | 0.023 |
|  | Sim(GPP+SWC) | 0.92*** | 0.090** | 0.18*** |  |
|  | Sim(CSIF) | 0.69*** | 0.20*** | 0.16*** |  |
|  | Sim(CSIF+SWC) | 0.88*** | 0.15*** | 0.18*** |  |
| US-SRM | Sim(GPP) | 0.94*** | 0.56*** | 0.52*** | 0.35*** |
|  | Sim(GPP+SWC) | 0.94*** | 0.61*** | 0.55*** |  |
|  | Sim(CSIF) | 0.77*** | 0.58*** | 0.53*** |  |
|  | Sim(CSIF+SWC) | 0.86*** | 0.58*** | 0.56*** |  |

\*\*\* Significant at the 0.001 level.
\*\* Significant at the 0.01 level.
\* Significant at the 0.05 level.

## 4. Discussion

4.1 Comparison of ET simulations with CSIF and EC GPP

The $G_c$-TSEB model using the OSM was tested at six dryland flux sites using either EC GPP or remote sensing CSIF product. EC GPP is regarded as the ideal indicator to constrain the $G_c$ model at the ecosystem level (De Kauwe et al., 2015; Medlyn et al., 2017). $G_c$-TSEB constrained by EC GPP was able to estimate half-hourly and daily ET well during both the growing and non-growing season of drylands. Thus, we used it as a benchmark to verify the capacity of a new global product, CSIF, to model ET in water-scarce regions.

The model performance using CSIF as a proxy for GPP showed high agreement with the model using EC GPP, particularly at the daily scale. CSIF captured the dynamics of vegetation responses to water availability providing valid biophysical signals for ET modeling over drylands. $G_c$-TSEB using CSIF showed superior capacity in very dry conditions (ES-BB) among ET modeling from previous studies relying on RS or field information without optimal values of green LAI and soil resistance (Table 3).

Table 3 Statistics from LE and H at six sites by various ET models from previous studies. Only simulations with CSIF in this study list in the table. If the model parameters were calibrated, the results of validation were selected for comparison. The unit (mm day$^{-1}$) of RMSE has been converted to W m$^{-2}$.

| Site | Source | LE R$^2$ | LE RMSE (W m$^{-2}$) | H R$^2$ | H RMSE (W m$^{-2}$) | ET Model type | Time interval | Forcing data |
|---|---|---|---|---|---|---|---|---|
| ES-BB | This study | 0.49 (0.79) | 33 (6) | 0.94 | 47 | G$_c$-TSEB | half-hourly (daily) | RS$^a$(CSIF) |
| | This study | 0.54 (0.79) | 32 (6) | 0.95 | 43 | G$_c$-TSEB | half-hourly (daily) | RS(CSIF)+SWC |
| | (Morillas et al., 2013a) | 0.39 | 105 | 0.77 | 64 | TSEB | 15-minute | RS (proximal) |
| | (Kustas et al., 2016) | 0.19 | 72 | 0.76 | 52 | TSEB | 15-minute | RS (proximal) |
| | (Li et al., 2019b) | | | | 54 | TSEB | 15-minute | RS (proximal) |
| | (Morillas et al., 2013b) | 0.47 | 6 | | | PML | daily | in-situ |
| | (Garcia et al., 2013) | 0.31 | 15 | | | PT-JPL | daily | RS (proximal) |
| | (Garcia et al., 2013) | 0.57 | 11 | | | PT-JPL | daily | in-situ |
| CN-Du2 | This study | 0.75 | 15 | | | G$_c$-TSEB | daily | RS(CSIF) |
| | This study | 0.81 | 12 | | | G$_c$-TSEB | daily | RS(CSIF)+SWC |
| | (Zhu et al., 2019) | 0.77 | 16 | | | SiTH | daily | in-situ |
| | (Xing et al., 2021) | 0.99 | 3 | | | PT-JPL | monthly | RS |
| | (Cao et al., 2021) | 0.71 | 16 | | | RS-PM | daily | RS |
| US-Var | This study | 0.90 | 15 | | | G$_c$-TSEB | daily | RS(CSIF) |
| | This study | 0.94 | 13 | | | G$_c$-TSEB | daily | RS(CSIF)+SWC |
| | (Dou and Yang, 2018) | 0.89 | 9 | | | Machine learning | daily | in-situ |
| | (Zhang et al., 2017) | 0.37 | 21 | | | PT-JPL | daily | in-situ |

| Site | Source | LE | | H | | ET Model type | Time interval | Forcing data |
|---|---|---|---|---|---|---|---|---|
| | | $R^2$ | RMSE (W m$^{-2}$) | $R^2$ | RMSE (W m$^{-2}$) | | | |
| | (Purdy et al., 2018) | 0.48 | 12 | | | PT-JPL | daily | RS |
| | (Holmes et al., 2018) | 0.25 | 86 | | | ALEXI | Weekly | RS |
| | (Wei and Zhu, 2021) | 0.49 | 29 | | | Surface temperature-vegetation index contextual model | daily | RS |
| | (Yebra et al., 2013) | 0.74 | 24 | | | PM-Gs | 16-day | in-situ |
| | (Chen et al., 2021b) | 0.95 | 3 | | | A deep learning HPM | monthly | in-situ |
| | (El Masri et al., 2019) | 0.62 | 13 | | | RS-PMPT | 8-day | in-situ |
| | (Bayat et al., 2019) | 0.95 | 11 | | | SCOPE-SM | daily | in-situ |
| | (Su et al., 2018) | 0.57 | 18 | | | PT-JPL combined with a hierarchical Bayesian (HB) method | monthly | in-situ |
| | (Li et al., 2019a) | 0.61 | 17 | | | PM | daily | in-situ |
| US-Wkg | This study | 0.76 | 13 | | | $G_c$-TSEB | daily | RS(CSIF) |
| | This study | 0.77 | 13 | | | $G_c$-TSEB | daily | RS(CSIF)+SWC |
| | (Purdy et al., 2018) | 0.96 | 24 | | | PT-JPL | daily | in-situ |
| | (Purdy et al., 2018) | 0.43 | 21 | | | PT-JPL | daily | RS |
| | (Zhang et al., 2017) | 0.47 | 16 | | | PT-JPL | daily | in-situ |
| | (Holmes et al., 2018) | 0.73 | 63 | | | ALEXI | weekly | RS |
| | (Wei and Zhu, 2021) | 0.50 | 28 | | | Surface temperature-vegetation index contextual model | daily | RS |
| | (Chen et al., 2021b) | 0.91 | 4 | | | A deep learning HPM | monthly | in-situ |
| | (El Masri et al., 2019) | 0.96 | 7 | | | RS-PMPT | 8-day | in-situ |
| | (Su et al., 2018) | 0.91 | 8 | | | PT-JPL combined with a hierarchical Bayesian (HB) method | monthly | in-situ |

| Site | Source | LE R² | LE RMSE (W m⁻²) | H R² | H RMSE (W m⁻²) | ET Model type | Time interval | Forcing data |
|---|---|---|---|---|---|---|---|---|
| | (Li et al., 2019a) | 0.74 | 12 | | | PM | daily | in-situ |
| | (Maheu et al., 2019) | 0.72 | 17 | | | HGS-MEP | half-hourly | in-situ |
| | (Zhang et al., 2019a) | 0.87 | 11 | | | PML-V2 | 8-day | RS |
| | (Mu et al., 2011) | 0.27 | 19 | | | RS-PM | daily | in-situ |
| | (Mu et al., 2011) | 0.26 | 20 | | | RS-PM | daily | RS |
| ES-LM2 | This study | 0.67 | 56 | 0.83 | 51 | G$_c$-TSEB | half-hourly | RS(CSIF) |
| | This study | 0.76 | 50 | 0.81 | 55 | G$_c$-TSEB | half-hourly | RS(CSIF)+SWC |
| | (Burchard-Levine et al., 2020) | 0.72 | 60 | 0.67 | 63 | TSEB-2S | half-hourly | RS (proximal) |
| | (Burchard-Levine et al., 2021b) | 0.56 | 57 | 0.64 | 40 | TSEB-2S | half-hourly | RS |
| US-SRM | This study | 0.74 (0.84) | 45 (12) | 0.81 | 61 | G$_c$-TSEB | half-hourly (daily) | RS(CSIF) |
| | This study | 0.80 (0.88) | 40 (11) | 0.83 | 57 | G$_c$-TSEB | half-hourly (daily) | RS(CSIF)+SWC |
| | (Ershadi et al., 2014) | 0.61 | 53 | | | SEBS | half-hourly | in-situ |
| | (Ershadi et al., 2014) | 0.48 | 112 | | | PM | half-hourly | in-situ |
| | (Ershadi et al., 2014) | 0.51 | 134 | | | Advection-aridity complementary method | half-hourly | in-situ |
| | (Ershadi et al., 2014) | 0.67 | 48 | | | PT-JPL | half-hourly | in-situ |
| | (Purdy et al., 2018) | 0.34 | 25 | | | PT-JPL | daily | RS |
| | (Zhang et al., 2017) | 0.21 | 15 | | | PT-JPL | daily | in-situ |
| | (Holmes et al., 2018) | 0.45 | 80 | | | ALEXI | Weekly | RS |

| Site | Source | LE | | H | | ET Model type | Time interval | Forcing data |
|---|---|---|---|---|---|---|---|---|
| | | $R^2$ | RMSE (W m$^{-2}$) | $R^2$ | RMSE (W m$^{-2}$) | | | |
| | (Wei and Zhu, 2021) | 0.42 | 34 | | | Surface temperature-vegetation index contextual model | daily | RS |
| | (Chen et al., 2021b) | 0.94 | 5 | | | A deep learning HPM | monthly | in-situ |
| | (Su et al., 2018) | 0.42 | 23 | | | PT-JPL combined with a hierarchical Bayesian (HB) method | monthly | in-situ |
| | (Zhang et al., 2019a) | 0.81 | 13 | | | PML-V2 | 8-day | RS |
| | (Li et al., 2019a) | 0.72 | 14 | | | PM | daily | in-situ |

[a] Remote sensing

However, simulations using CSIF were slightly inferior in quality to the ones using EC GPP at half-hourly time scales. Modeled LE using CSIF tend to be overestimated when LE is generally low and vice-versa (Fig. 7). However, at the diurnal scale, the deviations were not strong (Fig. 8). The model was run at a half-hourly scale while daily average CSIF was used to constrain GPP and therefore the canopy conductance in this study. However, SIF is driven by the variation of solar radiation, non-photochemical quenching (NPQ), dissipation of energy through heat, and environmental factors (air temperature, VPD), presenting strong diurnal and seasonal variations (Liu et al., 2019; Martini et al., 2021; Wang et al., 2021).

In TSEB-type models, energy (Burchard-Levine et al., 2020; Gan and Gao, 2015) and ET partitioning (Song et al., 2016a; Song et al., 2016b) are found to be sensitive to the error in the LST input. $G_c$-TSEB is quite different from the original TSEB (Kustas and Norman, 1999a) as well as N95 (Norman et al., 1995). The original TSEB uses RS LST as model input to achieve surface energy closure and predict heat fluxes (Kustas and Norman, 1999a). $T_c$ and $T_s$ are estimated from mono-angle LST. While in $G_c$-TSEB, LST is the model output rather than model input. Without LST, the solution to the energy balance equations requires parameterizing the turbulent processes between surface and atmosphere (Gan and Gao, 2015). $T_c$ and $T_s$ can be estimated iteratively by the surface energy balance equations to canopy and soil separately. This means that the accuracy of the LST simulation directly determines the reliability of the heat fluxes retrieval. Unsatisfactory results for LST (Fig. 10) and associated longwave net radiation (Fig. S1) simulation were observed at ES-LM2, both using EC GPP or CSIF in the model. As mentioned in Section 2.1, ES-LM2 is a typical tree-grass ecosystem with heterogeneous vegetation, consisting of low grassland (0.1-0.3 m) and high evergreen broadleaved trees (8 m). In our study, we assumed that there is a steady roughness parameter. This assumption made modeling simpler but inaccurate by ignoring the distinct canopy structures and phenology especially in tree-grass ecosystem (Burchard-Levine et al., 2021a; Burchard-Levine et al., 2020; Chen et al., 1999; Ryu et al., 2010).

4.2 Evaluation of ET partitioning compared to the uWUE method

In Section 3.5, we compared the ET partitioning by the $G_c$-TSEB model with the results by the uWUE method based on EC observations. There were high correlations between estimated T and $T_{uWUE}$, particularly when using root-zone SWC (5cm-40cm in this study). The $G_c$-TSEB model was highly sensitive to the inclusion (or lack thereof) of root-zone SWC during germination or greening (depending on life form) and during the senescence of vegetation (Fig. 12). However, most natural drylands are limited by soil moisture supply from precipitation, but the stress level differs between crops and forests. The former depends on regular irrigation and the latter on their deep roots, potentially accessing groundwater (Scott et al., 2014). Thus, accounting for the effect of soil moisture on stomatal regulation significantly improve ET modeling in water-limited regions (Brust et al., 2021; Purdy et al., 2018; Stocker et al., 2018). Besides, SIF-based estimation of T in this study relied on the tight relationship between GPP and SIF. However, SIF is the byproduct of light reaction of photosynthesis rather than direct $CO_2$ assimilation (Magney et al., 2020), and also dependent on the dissipation of excess of light through NPQ (Martini et al., 2021; Wang et al., 2020). Thus, the indirect link of SIF-GPP makes the SIF-T relationship rather complicated, particularly under environmental stress conditions, such as water limitation (Damm et al., 2021; Shan et al., 2021). The root-zone SWC constrained the canopy conductance, thus improving the T and ET simulation, compared with not including

SWC as a predictor in the model. The improvements indicated that it is feasible to apply root-zone SWC regulation in SIF-based ET modeling under water stress.

There were some differences between E from $G_c$-TSEB and $E_{uWUE}$, though the trends of the two curves were consistent at most sites. During the rainy season, E estimated by $G_c$-TSEB was slightly lower than the observational one based on the uWUE partitioning method, particularly at ES-BB, US-Var, US-Wkg, ES-LM2. As the soil evaporation can increase rapidly following a precipitation event in drylands, E was likely underestimated by the model, which may be the primary reason for the model systematically underestimation of peak ET (Fig. 5 and Fig. 7). Worth noting, in the dry season when almost no precipitation occurred, $E_{uWUE}$ still had high values at some sites, such as US-Var (Day 145-161 in 2010) and ES-LM2 (Day 137-241 in 2016), while the E estimated by $G_c$-TSEB fell to very small values, close to zero. The sparse vegetation cover and different phenology of various species (ES-LM2) in drylands might also challenge the assumption that $uWUE_p$ is a constant and result in errors of the uWUE method (Xu et al., 2021; Zhou et al., 2016). E is regulated directly by precipitation in $G_c$-TSEB, which seems more reasonable for drylands.

4.3 Uncertainties and implications

Although the $G_c$-TSEB model based on an OSM and CSIF showed good performance in drylands, there are some shortcomings to be noted. The spatial resolution of CSIF (0.05°) is coarser than the footprint of EC tower (less than 1 km). Taking US-Var (GRA) as an example, the grass withered in summer because of heat and water scarcity, as shown the seasonal cycle of the EC GPP (Fig. 2). However, the SIF still remained high during the drought, particularly in the summer of 2011, finally leading to an overestimation of ET (Fig. 8). This is most likely due to the preponderance of forests in the southwest of the target pixel (Fig. 1), whose photosynthetic activity would also affect the SIF signals. Large pixel sizes cannot descript the heterogeneity within a pixel, particularly in tree-grass ecosystems (Burchard-Levine et al., 2021b). Moreover, the sparse canopy in drylands can also result in a low ratio of signal to noise for satellites (Smith et al., 2019; Yan et al., 2019).

There are fitting parameters that need to be calibrated in the model base on *in-situ* data, especially $g_1$, $k_1$, and $\Omega$ that were found to have a large influence on the simulation (Fig. 11). In this study, we fixed the parameters during the entire period. However, the empirical stomatal slope $g_1$ critically depends on climatic characteristics (Migliavacca et al., 2021). The canopy structure (Zhang et al., 2016), and seasonal phenology (Chen et al., 2021a) also have an impact on the linear regression slope, $k_1$, between SIF and EC GPP. The parameter $\Omega$ related to under-canopy resistance could strongly affect the fluxes estimation in relatively drier sites (Section 3.4). Kustas et al. (2016) pointed that the rocky soil surface and clumped canopy are sources of large biases in LE and H estimation by TSEB (Morillas et al., 2013a). Hence, it is essential to consider the properties of canopy structure and near-surface turbulence in ET modeling for drylands vegetation (Li et al., 2019b; Villagarcía et al., 2007). To overcome the spatial limitation of EC observations, The CSIF version of $G_c$-TSEB could also be calibrated by LST detected by infrared thermal sensors or data assimilation (e.g. Kalman filter), which has proved to be practical in a series of studies (Gan and Gao, 2015; Gan et al., 2019).

To fit the CSIF pixels (0.05°), we upscaled the OSM from leaf to canopy level by replacing the vapor pressure deficit at the leaf surface ($D_s$) with the atmospheric VPD, in both EC GPP and CSIF scenarios. This assumes that there is no difference in temperature between the leaf and air temperatures (Feng et al., 2021; Shan et al., 2021). Yet, a large fraction of the available energy dissipates as sensible heat in water-limited regions, resulting in large imbalance between leaf and air temperatures. Therefore, this simple assumption might bring some biases in the ET estimation in drylands.

## 5. Conclusions

In this study, we applied a remote sensing proxy of gross primary production (GPP), the Contiguous Solar-Induced Chlorophyll Fluorescence (CSIF), to constrain an optimal stomatal model with a two-source energy balance model over six dryland sites, and comparing the results to using in-situ GPP instead of CSIF as a benchmark. The following conclusions can be drawn:

(1) The model using the in-situ GPP performed well in drylands, with half-hourly and daily average determination coefficient of 0.73 and 0.86 respectively.

(2) The simulation including CSIF showed similar performance to the one using in-situ GPP, with determination coefficient of 0.79 and root-mean-square error of 0.46 mm day$^{-1}$. The results improved the remote sensing-based evapotranspiration simulations at some of the drier sites.

(3) The transpiration simulated by the model, based on either in-situ GPP or CSIF, was in good agreement with the observational timeseries, derived by the underlying water use efficiency method. The crucial role of also including soil water content in the model, in addition to GPP, was especially important on evapotranspiration during the senescence period.

Our results show the capacity of CSIF to constraint regional estimates of transpiration specifically under very limited evapotranspiration conditions, with great potential of remote sensed solar-induced chlorophyll fluorescence on regional evapotranspiration estimation. Our study also demonstrates the power of blending two modeling approaches that are typically separated: latent heat focusing on stomatal conductance and the two-source model focusing on soil and canopy sensible heat fluxes.

## Credit Author Statement

Jingyi Bu: Methodology, Software, Writing-original draft, Writing-review & editing. Guojing Gan: Methodology, Writing-review & editing. Jiahao Chen: Software, Writing-review & editing. Yanxin Su: Writing-review & editing. Mengjia Yuan: Writing-review & editing. Yanchun Gao: Conceptualization, Writing-review & editing, Supervision, Funding acquisition, Project administration. Francisco Domingo: Resources, Writing-review & editing. Mirco Migliavacca: Resources, Writing-review & editing. Tarek S. El-Madany: Resources, Writing-review & editing. Pierre Gentine: Resources, Writing-review & editing. Mónica García: Conceptualization, Methodology, Writing-original draft, Writing-review & editing, Supervision.

## Acknowledgments

This research was jointly funded by the National Science Foundation of China (42071054), the National Natural Science Foundation of China (41430861), the EU and Innovation Fund Denmark (IFD) for funding the FORWARD collaborative international consortium (ERA-NET co-fund WaterWorks2015- Water JPI initiative); Danida Fellowship Centre (EOForChina project, file number: 18-M01-DTU); Innovation Fund Denmark (ChinaWaterSense project, file number: 8087-00002B) and from the INTEGRATYON3 project (PID2020-117825GB-C21 and C22) through the Spanish Ministry of Economy and Competitiveness and projects BAGAMET (P20_00016) and Adquisición de Equipamiento científico-tecnico (IE17_5560_EEZA), funded by the Counseling of Economy, Innovation, Science and Employment from the Government of Andalucía, all including European Union ERDF funds, and the China Scholarship Council. This work used meteorological and eddy covariance data at CN-Du2, US-Var, US-Wkg, and US-SRM acquired and shared by the FLUXNET community (http://fluxnet.fluxdata.org/data/fluxnet2015-dataset/). Gentine acknowledges funding from NASA grant 80NSSC20K1792 P00002 and NNX16AO16H S004. The contiguous solar-induced fluorescence (CSIF) datasets were downloaded from the OSF CSIF repository (https://osf.io/8xqy6/) developed by Yao Zhang and colleagues. MM and TEM acknowledge the support of the Alexander von Humboldt Foundation and the 2013 Max Planck Research Prize to Markus Reichstein for funding the research site ES-LM2.

# References

Anderson, M.C., Norman, J.M., Mecikalski, J.R., Otkin, J.A. and Kustas, W.P., 2007. A climatological study of evapotranspiration and moisture stress across the continental United States based on thermal remote sensing: 2. Surface moisture climatology. Journal of Geophysical Research, 112(D11).

Baldocchi, D.D., Xu, L. and Kiang, N., 2004. How plant functional-type, weather, seasonal drought, and soil physical properties alter water and energy fluxes of an oak–grass savanna and an annual grassland. Agricultural and Forest Meteorology, 123(1-2): 13-39.

Ball, J.T., Woodrow, I.E. and Berry, J.A., 1987. A model predicting stomatal conductance and its contribution to the control of photosynthesis under different environmental conditions, Progress in photosynthesis research. Springer, pp. 221-224.

Bayat, B., van der Tol, C., Yang, P.Q. and Verhoef, W., 2019. Extending the SCOPE model to combine optical reflectance and soil moisture observations for remote sensing of ecosystem functioning under water stress conditions. Remote Sensing of Environment, 221: 286-301.

Berdugo, M., Vidiella, B., Solé, R.V. and Maestre, F.T., 2021. Ecological mechanisms underlying aridity thresholds in global drylands. Functional Ecology.

Berg, A. and McColl, K.A., 2021. No projected global drylands expansion under greenhouse warming. Nature Climate Change, 11(4): 331-337.

Bonan, G.B., Williams, M., Fisher, R.A. and Oleson, K.W., 2014. Modeling stomatal conductance in the earth system: linking leaf water-use efficiency and water transport along the soil–plant–atmosphere continuum. Geoscientific Model Development, 7(5): 2193-2222.

Brust, C. et al., 2021. Using SMAP Level-4 soil moisture to constrain MOD16 evapotranspiration over the contiguous USA. Remote Sensing of Environment, 255: 112277.

Bu, J. et al., 2021. Biophysical constraints on evapotranspiration partitioning for a conductance-based two source energy balance model. Journal of Hydrology, 603: 127179.

Burchard-Levine, V. et al., 2021a. A remote sensing-based three-source energy balance model to improve global estimations of evapotranspiration in semi-arid tree-grass ecosystems. Glob Chang Biol.

Burchard-Levine, V. et al., 2021b. The effect of pixel heterogeneity for remote sensing based retrievals of evapotranspiration in a semi-arid tree-grass ecosystem. Remote Sensing of Environment, 260: 112440.

Burchard-Levine, V. et al., 2020. Seasonal Adaptation of the Thermal-Based Two-Source Energy Balance Model for Estimating Evapotranspiration in a Semiarid Tree-Grass Ecosystem. Remote Sensing, 12(6): 904.

Cao, M. et al., 2021. Multiple sources of uncertainties in satellite retrieval of terrestrial actual evapotranspiration. Journal of Hydrology, 601: 126642.

Cavanaugh, M.L., Kurc, S.A. and Scott, R.L., 2011. Evapotranspiration partitioning in semiarid shrubland ecosystems: a two-site evaluation of soil moisture control on transpiration. Ecohydrology, 4(5): 671-681.

Chen, A. et al., 2021a. Seasonal changes in GPP/SIF ratios and their climatic determinants across the Northern Hemisphere. Glob Chang Biol, 27(20): 5186-5197.

Chen, J., Dafflon, B., Tran, A.P., Falco, N. and Hubbard, S.S., 2021b. A deep learning hybrid predictive modeling (HPM) approach for estimating evapotranspiration and ecosystem respiration. Hydrology and Earth System Sciences, 25(11): 6041-6066.

Chen, J., Liu, J., Cihlar, J. and Goulden, M., 1999. Daily canopy photosynthesis model through temporal and spatial scaling for remote sensing applications. Ecological modelling, 124(2-3): 99-119.

Chen, S. et al., 2009. Energy balance and partition in Inner Mongolia steppe ecosystems with different land use types. Agricultural and Forest Meteorology, 149(11): 1800-1809.

D'Odorico, P., Porporato, A. and Runyan, C.W., 2019. Dryland ecohydrology. Springer International Publishing.

Damm, A., Haghighi, E., Paul-Limoges, E. and van der Tol, C., 2021. On the seasonal relation of sun-induced chlorophyll fluorescence and transpiration in a temperate mixed forest. Agricultural and Forest Meteorology, 304-305: 108386.

De Kauwe, M.G. et al., 2015. A test of an optimal stomatal conductance scheme within the CABLE land surface model. Geoscientific Model Development, 8(2): 431-452.

Dou, X. and Yang, Y., 2018. Evapotranspiration estimation using four different machine learning approaches in different terrestrial ecosystems. Computers and Electronics in Agriculture, 148: 95-106.

Du, S. et al., 2018. Retrieval of global terrestrial solar-induced chlorophyll fluorescence from TanSat satellite. Science Bulletin, 63(22): 1502-1512.

El-Madany, T.S. et al., 2018. Drivers of spatio-temporal variability of carbon dioxide and energy fluxes in a Mediterranean savanna ecosystem. Agricultural and Forest Meteorology, 262: 258-278.

El-Madany, T.S. et al., 2021. How Nitrogen and Phosphorus Availability Change Water Use Efficiency in a Mediterranean Savanna Ecosystem. Journal of Geophysical Research: Biogeosciences, 126(5).

El Masri, B., Rahman, A.F. and Dragoni, D., 2019. Evaluating a new algorithm for satellite-based evapotranspiration for North American ecosystems: Model development and validation. Agricultural and Forest Meteorology, 268: 234-248.

Ershadi, A., McCabe, M.F., Evans, J.P., Chaney, N.W. and Wood, E.F., 2014. Multi-site evaluation of terrestrial evaporation models using FLUXNET data. Agricultural and Forest Meteorology, 187: 46-61.

Feng, H. et al., 2021. Modeling Transpiration with Sun-Induced Chlorophyll Fluorescence Observations via

Carbon-Water Coupling Methods. Remote Sensing, 13(4): 804.

Fensholt, R., Sandholt, I. and Rasmussen, M.S., 2004. Evaluation of MODIS LAI, fAPAR and the relation between fAPAR and NDVI in a semi-arid environment using in situ measurements. Remote Sensing of Environment, 91(3-4): 490-507.

Frankenberg, C. et al., 2011. New global observations of the terrestrial carbon cycle from GOSAT: Patterns of plant fluorescence with gross primary productivity. Geophysical Research Letters, 38(17): n/a-n/a.

Frankenberg, C. et al., 2014. Prospects for chlorophyll fluorescence remote sensing from the Orbiting Carbon Observatory-2. Remote Sensing of Environment, 147: 1-12.

Gan, G. and Gao, Y., 2015. Estimating time series of land surface energy fluxes using optimized two source energy balance schemes: Model formulation, calibration, and validation. Agricultural and Forest Meteorology, 208: 62-75.

Gan, G.J. et al., 2019. An optimized two source energy balance model based on complementary concept and canopy conductance. Remote Sensing of Environment, 223: 243-256.

Gao, Y., Gan, G., Liu, M. and Wang, J., 2016. Evaluating soil evaporation parameterizations at near-instantaneous scales using surface dryness indices. Journal of Hydrology, 541: 1199-1211.

Garcia, M. et al., 2008. Monitoring land degradation risk using ASTER data: The non-evaporative fraction as an indicator of ecosystem function. Remote Sensing of Environment, 112(9): 3720-3736.

Garcia, M. et al., 2013. Actual evapotranspiration in drylands derived from in-situ and satellite data: Assessing biophysical constraints. Remote Sensing of Environment, 131: 103-118.

Garcia, M. and Ustin, S.L., 2001. Detection of interannual vegetation responses to climatic variability using AVIRIS data in a coastal savanna in California. Ieee Transactions on Geoscience and Remote Sensing, 39(7): 1480-1490.

Gentine, P. et al., 2019. Coupling between the terrestrial carbon and water cycles-a review. Environmental Research Letters, 14(8).

Granger, R.J. and Gray, D., 1989. Evaporation from natural nonsaturated surfaces. Journal of Hydrology, 111(1-4): 21-29.

Guanter, L. et al., 2014. Global and time-resolved monitoring of crop photosynthesis with chlorophyll fluorescence. Proc Natl Acad Sci U S A, 111(14): E1327-33.

He, T., Gao, F., Liang, S. and Peng, Y., 2019. Mapping Climatological Bare Soil Albedos over the Contiguous United States Using MODIS Data. Remote Sensing, 11(6): 666.

Holmes, T.R., Hain, C., Crow, W.T., Anderson, M.C. and Kustas, W.P., 2018. Microwave implementation of two-source energy balance approach for estimating evapotranspiration. Hydrol Earth Syst Sci, 22(2): 1351-1369.

Hu, Z. et al., 2013. Modeling evapotranspiration by combing a two-source model, a leaf stomatal model, and a light-use efficiency model. Journal of Hydrology, 501: 186-192.

Hu, Z. et al., 2017. Modeling and partitioning of regional evapotranspiration using a satellite-driven water-carbon coupling model. Remote Sensing, 9(1).

Jarvis, P.G., 1976. The Interpretation of the Variations in Leaf Water Potential and Stomatal Conductance Found in Canopies in the Field. Philosophical Transactions of the Royal Society B: Biological Sciences, 273(927): 593-610.

Joiner, J. et al., 2013. Global monitoring of terrestrial chlorophyll fluorescence from moderate-spectral-resolution near-infrared satellite measurements: methodology, simulations, and application to GOME-2. Atmospheric Measurement Techniques, 6(10): 2803-2823.

Koehler, P. et al., 2018. Global retrievals of solar induced chlorophyll fluorescence with TROPOMI: first results and inter-sensor comparison to OCO-2. Geophys Res Lett, 45(19): 10456-10463.

Kool, D., Kustas, W.P., Ben-Gal, A. and Agam, N., 2021. Energy partitioning between plant canopy and soil, performance of the two-source energy balance model in a vineyard. Agricultural and Forest Meteorology, 300: 108328.

Kustas, W.P. et al., 2016. Revisiting the paper "Using radiometric surface temperature for surface energy flux estimation in Mediterranean drylands from a two-source perspective". Remote Sensing of Environment, 184: 645-653.

Kustas, W.P. and Norman, J.M., 1999a. Evaluation of soil and vegetation heat flux predictions using a simple two-source model with radiometric temperatures for partial canopy cover. Agricultural and Forest Meteorology, 94(1): 13-29.

Kustas, W.P. and Norman, J.M., 1999b. Reply to comments about the basic equations of dual-source vegetation–atmosphere transfer models. Agricultural and Forest Meteorology, 94(3-4): 275-278.

Kustas, W.P., Zhan, X. and Schmugge, T.J., 1998. Combining optical and microwave remote sensing for mapping energy fluxes in a semiarid watershed. Remote Sensing of Environment, 64(2): 116-131.

Leuning, R., Zhang, Y.Q., Rajaud, A., Cleugh, H. and Tu, K., 2008. A simple surface conductance model to estimate regional evaporation using MODIS leaf area index and the Penman-Monteith equation. Water Resources Research, 44(10).

Li, J. et al., 2019a. An Algorithm Differentiating Sunlit and Shaded Leaves for Improving Canopy Conductance and Vapotranspiration Estimates. Journal of Geophysical Research: Biogeosciences, 124(4): 807-824.

Li, S. et al., 2015. Comparison of several surface resistance models for estimating crop evapotranspiration over the entire growing season in arid regions. Agricultural and Forest Meteorology, 208: 1-15.

Li, X. et al., 2018. Solar-induced chlorophyll fluorescence is strongly correlated with terrestrial photosynthesis for a wide variety of biomes: First global analysis based on OCO-2 and flux tower observations. Glob Chang Biol, 24(9): 3990-4008.

Li, Y. et al., 2019b. Evaluating Soil Resistance Formulations in Thermal-Based Two-Source Energy Balance (TSEB) Model: Implications for Heterogeneous Semiarid and Arid Regions. Water Resources Research, 55(2): 1059-1078.

Liu, L., Guan, L. and Liu, X., 2017. Directly estimating diurnal changes in GPP for C3 and C4 crops using far-red sun-induced chlorophyll fluorescence. Agricultural and Forest Meteorology, 232: 1-9.

Liu, L. et al., 2020. Soil moisture dominates dryness stress on ecosystem production globally. Nat Commun, 11(1): 4892.

Liu, L. et al., 2019. The Impacts of Growth and Environmental Parameters on Solar-Induced Chlorophyll Fluorescence at Seasonal and Diurnal Scales. Remote Sensing, 11(17): 2002.

López-Ballesteros, A. et al., 2017. Subterranean ventilation of allochthonous CO 2 governs net CO 2 exchange in a semiarid Mediterranean grassland. Agricultural and Forest Meteorology, 234-235: 115-126.

Lu, X. et al., 2018. Potential of solar-induced chlorophyll fluorescence to estimate transpiration in a temperate forest. Agricultural and Forest Meteorology, 252: 75-87.

Ma, S., Baldocchi, D.D., Xu, L. and Hehn, T., 2007. Inter-annual variability in carbon dioxide exchange of an oak/grass savanna and open grassland in California. Agricultural and Forest Meteorology, 147(3): 157-171.

Maes, W.H. et al., 2020. Sun-induced fluorescence closely linked to ecosystem transpiration as evidenced by satellite data and radiative transfer models. Remote Sensing of Environment, 249: 112030.

Magney, T.S., Barnes, M.L. and Yang, X., 2020. On the Covariation of Chlorophyll Fluorescence and Photosynthesis Across Scales. Geophysical Research Letters, 47(23).

Maheu, A., Hajji, I., Anctil, F., Nadeau, D.F. and Therrien, R., 2019. Using the maximum entropy production approach to integrate energy budget modelling in a hydrological model. Hydrology and Earth System Sciences, 23(9): 3843-3863.

Martini, D. et al., 2021. Heatwave breaks down the linearity between sun-induced fluorescence and gross primary production. New Phytol.

Medlyn, B.E. et al., 2017. How do leaf and ecosystem measures of water-use efficiency compare? New Phytol, 216(3): 758-770.

Medlyn, B.E. et al., 2011. Reconciling the optimal and empirical approaches to modelling stomatal conductance. Global Change Biology, 17(6): 2134-2144.

Meng, F., Huang, L., Chen, A., Zhang, Y. and Piao, S., 2021. Spring and autumn phenology across the Tibetan Plateau inferred from normalized difference vegetation index and solar-induced chlorophyll fluorescence. Big Earth Data, 5(2): 182-200.

Miao, H. et al., 2009. Cultivation and grazing altered evapotranspiration and dynamics in Inner Mongolia steppes. Agricultural and Forest Meteorology, 149(11): 1810-1819.

Migliavacca, M. et al., 2021. The three major axes of terrestrial ecosystem function. Nature, 598(7881): 468-472.

Mirzabaev, A. et al., 2019. Desertification.

Mohammed, G.H. et al., 2019. Remote sensing of solar-induced chlorophyll fluorescence (SIF) in vegetation: 50 years of progress. Remote Sensing of Environment, 231.

Morillas, L. et al., 2013a. Using radiometric surface temperature for surface energy flux estimation in Mediterranean drylands from a two-source perspective. Remote Sensing of Environment, 136: 234-246.

Morillas, L. et al., 2013b. Improving evapotranspiration estimates in Mediterranean drylands: The role of soil evaporation. Water Resources Research, 49(10): 6572-6586.

Mu, M. et al., 2021. Evaluating a land surface model at a water-limited site: implications for land surface contributions to droughts and heatwaves. Hydrology and Earth System Sciences, 25(1): 447-471.

Mu, Q., Zhao, M. and Running, S.W., 2011. Improvements to a MODIS global terrestrial evapotranspiration algorithm. Remote Sensing of Environment, 115(8): 1781-1800.

Myneni, R., Knyazikhin, Y. and Park, T., 2015. MOD15A2H MODIS/terra leaf area index/FPAR 8-Day L4 global 500m SIN grid V006. NASA EOSDIS Land Processes DAAC.

Norman, J.M., Kustas, W.P. and Humes, K.S., 1995. Source approach for estimating soil and vegetation energy fluxes in observations of directional radiometric surface-temperature. Agricultural and Forest Meteorology, 77(3-4): 263-293.

Nossent, J., Elsen, P. and Bauwens, W., 2011. Sobol' sensitivity analysis of a complex environmental model. Environmental Modelling & Software, 26(12): 1515-1525.

Pagán, B., Maes, W., Gentine, P., Martens, B. and Miralles, D., 2019. Exploring the Potential of Satellite Solar-Induced Fluorescence to Constrain Global Transpiration Estimates. Remote Sensing, 11(4).

Porporato, A., Laio, F., Ridolfi, L. and Rodriguez-Iturbe, I., 2001. Plants in water-controlled ecosystems: active role in hydrologic processes and response to water stress: III. Vegetation water stress. Advances in Water Resources, 24(7): 725-744.

Purdy, A.J. et al., 2018. SMAP soil moisture improves global evapotranspiration. Remote Sensing of

Environment, 219: 1-14.

Reynolds, J.F. et al., 2007. Global desertification: building a science for dryland development. Science, 316(5826): 847-851.

Ryu, Y. et al., 2011. Integration of MODIS land and atmosphere products with a coupled-process model to estimate gross primary productivity and evapotranspiration from 1 km to global scales. Global Biogeochemical Cycles, 25(4): n/a-n/a.

Ryu, Y. et al., 2010. How to quantify tree leaf area index in an open savanna ecosystem: A multi-instrument and multi-model approach. Agricultural and Forest Meteorology, 150(1): 63-76.

Savitzky, A. and Golay, M.J., 1964. Smoothing and differentiation of data by simplified least squares procedures. Analytical chemistry, 36(8): 1627-1639.

Schaaf, C. and Wang, Z., 2015. MCD43A1 MODIS/Terra+ Aqua BRDF/Albedo Model Parameters Daily L3 Global–500m V006, NASA EOSDIS Land Processes DAAC.

Schlesinger, W.H. and Jasechko, S., 2014. Transpiration in the global water cycle. Agricultural and Forest Meteorology, 189-190: 115-117.

Scott, R.L., 2010. Using watershed water balance to evaluate the accuracy of eddy covariance evaporation measurements for three semiarid ecosystems. Agricultural and Forest Meteorology, 150(2): 219-225.

Scott, R.L., Hamerlynck, E.P., Jenerette, G.D., Moran, M.S. and Barron-Gafford, G.A., 2010. Carbon dioxide exchange in a semidesert grassland through drought-induced vegetation change. Journal of Geophysical Research, 115(G3).

Scott, R.L. et al., 2014. When vegetation change alters ecosystem water availability. Glob Chang Biol, 20(7): 2198-210.

Shan, N. et al., 2019. Modeling canopy conductance and transpiration from solar-induced chlorophyll fluorescence. Agricultural and Forest Meteorology, 268: 189-201.

Shan, N. et al., 2021. A model for estimating transpiration from remotely sensed solar-induced chlorophyll fluorescence. Remote Sensing of Environment, 252.

Smith, W.K. et al., 2018. Chlorophyll Fluorescence Better Captures Seasonal and Interannual Gross Primary Productivity Dynamics Across Dryland Ecosystems of Southwestern North America. Geophysical Research Letters, 45(2): 748-757.

Smith, W.K. et al., 2019. Remote sensing of dryland ecosystem structure and function: Progress, challenges, and opportunities. Remote Sensing of Environment, 233: 111401.

Sobol', I.M., 2001. Global sensitivity indices for nonlinear mathematical models and their Monte Carlo estimates. Mathematics and Computers in Simulation, 55(1): 271-280.

Song, L. et al., 2016a. Applications of a thermal-based two-source energy balance model using Priestley-Taylor approach for surface temperature partitioning under advective conditions. Journal of Hydrology, 540: 574-587.

Song, L.S. et al., 2016b. Application of remote sensing-based two-source energy balance model for mapping field surface fluxes with composite and component surface temperatures. Agricultural and Forest Meteorology, 230: 8-19.

Stocker, B.D. et al., 2018. Quantifying soil moisture impacts on light use efficiency across biomes. New Phytol, 218(4): 1430-1449.

Su, Y. et al., 2018. A hierarchical Bayesian approach for multi-site optimization of a satellite-based evapotranspiration model. Hydrological Processes, 32(26): 3907-3923.

Sugita, M. and Brutsaert, W., 1993. Cloud effect in the estimation of instantaneous downward longwave radiation. Water Resources Research, 29(3): 599–605.

Sun, X., Wilcox, B.P. and Zou, C.B., 2019. Evapotranspiration partitioning in dryland ecosystems: A global meta-analysis of in situ studies. Journal of Hydrology, 576: 123-136.

Swift, J.E.M.C.D., 1993. Climate variability, ecosystem stability and the implications for range and livestock development. Range ecology at disequilibrium Overseas Development Institute London: 31-41.

Twine, T.E. et al., 2000. Correcting eddy-covariance flux underestimates over a grassland. Agricultural and forest meteorology, 103(3): 279-300.

Villagarcía, L., Were, A., Domingo, F., García, M. and Alados-Arboledas, L., 2007. Estimation of soil boundary-layer resistance in sparse semiarid stands for evapotranspiration modelling. Journal of Hydrology, 342(1): 173-183.

Wang, L. et al., 2012. Dryland ecohydrology and climate change: critical issues and technical advances. Hydrology and Earth System Sciences, 16(8): 2585-2603.

Wang, N. et al., 2021. Diurnal variation of sun-induced chlorophyll fluorescence of agricultural crops observed from a point-based spectrometer on a UAV. International Journal of Applied Earth Observation and Geoinformation, 96: 102276.

Wang, X., Chen, J.M. and Ju, W., 2020. Photochemical reflectance index (PRI) can be used to improve the relationship between gross primary productivity (GPP) and sun-induced chlorophyll fluorescence (SIF). Remote Sensing of Environment, 246: 111888.

Wei, J. and Zhu, W., 2021. An operational parameterization scheme of surface temperature-vegetation index contextual model for large-scale temporally continuous evapotranspiration estimation: The case study of contiguous United States. Journal of Hydrology, 602: 126805.

Xing, W., Wang, W., Shao, Q., Song, L. and Cao, M., 2021. Estimation of Evapotranspiration and Its Components across China Based on a Modified Priestley–Taylor Algorithm Using Monthly Multi-Layer Soil Moisture Data. Remote Sensing, 13(16): 3118.

Xu, Z. et al., 2021. Evapotranspiration partitioning for multiple ecosystems within a dryland watershed: Seasonal variations and controlling factors. Journal of Hydrology, 598: 126483.

Yan, D., Scott, R.L., Moore, D.J.P., Biederman, J.A. and Smith, W.K., 2019. Understanding the relationship between vegetation greenness and productivity across dryland ecosystems through the integration of PhenoCam, satellite, and eddy covariance data. Remote Sensing of Environment, 223: 50-62.

Yebra, M., Van Dijk, A., Leuning, R., Huete, A. and Guerschman, J.P., 2013. Evaluation of optical remote sensing to estimate actual evapotranspiration and canopy conductance. Remote Sensing of Environment, 129: 250-261.

Zeng, X. et al., 2005. Treatment of undercanopy turbulence in land models. J Climate, 18(23): 5086-5094.

Zhang, K. et al., 2017. Parameter sensitivity analysis and optimization for a satellite-based evapotranspiration model across multiple sites using Moderate Resolution Imaging Spectroradiometer and flux data. Journal of Geophysical Research: Atmospheres, 122(1): 230-245.

Zhang, Y., Commane, R., Zhou, S., Williams, A.P. and Gentine, P., 2020. Light limitation regulates the response of autumn terrestrial carbon uptake to warming. Nature Climate Change, 10(8): 739-743.

Zhang, Y. et al., 2016. Model-based analysis of the relationship between sun-induced chlorophyll fluorescence and gross primary production for remote sensing applications. Remote Sensing of Environment, 187: 145-155.

Zhang, Y., Joiner, J., Alemohammad, S.H., Zhou, S. and Gentine, P., 2018a. A global spatially contiguous solar-induced fluorescence (CSIF) dataset using neural networks. Biogeosciences, 15(19): 5779-

5800.

Zhang, Y. et al., 2019a. Coupled estimation of 500 m and 8-day resolution global evapotranspiration and gross primary production in 2002–2017. Remote Sensing of Environment, 222: 165-182.

Zhang, Y. et al., 2010. Using long-term water balances to parameterize surface conductances and calculate evaporation at 0.05 spatial resolution. Water Resources Research, 46(5).

Zhang, Y. et al., 2018b. On the relationship between sub-daily instantaneous and daily total gross primary production: Implications for interpreting satellite-based SIF retrievals. Remote Sensing of Environment, 205: 276-289.

Zhang, Z., Chen, J.M., Guanter, L., He, L. and Zhang, Y., 2019b. From Canopy-Leaving to Total Canopy Far-Red Fluorescence Emission for Remote Sensing of Photosynthesis: First Results From TROPOMI. Geophysical Research Letters, 46(21): 12030-12040.

Zhou, K., Zhang, Q., Xiong, L. and Gentine, P., 2022. Estimating evapotranspiration using remotely sensed solar-induced fluorescence measurements. Agricultural and Forest Meteorology, 314: 108800.

Zhou, S., Yu, B., Zhang, Y., Huang, Y. and Wang, G., 2016. Partitioning evapotranspiration based on the concept of underlying water use efficiency. Water Resources Research, 52(2): 1160-1175.

Zhu, G. et al., 2019. Development and evaluation of a simple hydrologically based model for terrestrial evapotranspiration simulations. Journal of Hydrology, 577: 123928.